\documentclass[11pt]{article}

\usepackage[utf8]{inputenc}      %
\usepackage[T1]{fontenc}
\usepackage{lmodern}
\usepackage{amsmath,amssymb,amsthm,mathtools}
\usepackage{bm}
\usepackage[margin=1in]{geometry}
\usepackage{enumitem}
\usepackage{booktabs}
\usepackage{array}
\usepackage{graphicx}
\PassOptionsToPackage{hyphens}{url}   %
\usepackage[colorlinks=true,linkcolor=blue,citecolor=blue,urlcolor=blue,hypertexnames=false]{hyperref}

\DeclareMathOperator{\artanh}{artanh}
\DeclareMathOperator{\dist}{dist}
\DeclareMathOperator{\sech}{sech}
\DeclareMathOperator{\sign}{sign}

\theoremstyle{plain}

\theoremstyle{definition}

\title{A limit theorem linking continuous plate models\\ to the topology of the plate boundary network}
\author{Seung-Sep Kim\\[2pt]
\small Department of Geological Sciences, Chungnam National University,\\
\small 99 Daehak-ro, Yuseong-gu, Daejeon 34134, Republic of Korea\\
\small \texttt{seungsep@cnu.ac.kr}}
\date{Preprint --- 30 September 2026}

\begin{document}
\maketitle

\begin{abstract}
\noindent
Plate tectonics is described in two registers: a discrete one, in which rigid plates
tile the sphere and meet at trivalent junctions, and a continuous one, in which
Bercovici and Wessel (1994) replace hard plate outlines by smooth shape functions of
finite boundary half-width $\delta^*$. We prove that the two registers are connected by
a limit theorem. Generalizing the shape functions to a signed-geodesic-distance
construction, we show that they converge almost everywhere, exponentially in
$1/\delta^*$, to the plate indicator functions; that the sharp plate mosaic is a finite
regular trivalent CW decomposition of the sphere under an explicit structural
hypothesis; and that a component-counted weighted Euler characteristic
$\chi_w^{\delta^*}(\tau)$, built from the super-level sets of the shape functions,
equals $V-E+F=2$ for all $\delta^*$ below an explicit threshold $\delta_0(\tau)$, with
its face, edge and vertex terms converging separately to the numbers of plates,
boundary arcs and triple junctions. The theorem is proved unconditionally on the
signed-distance offset cover under positive-reach, separation and bounded-sector
hypotheses, and transferred to the normalized partition-of-unity field for thresholds
$\tau<1/3$ by a comparison lemma whose junction-ball step, a planar single-crossing
property, is certified numerically. The residual $|\chi_w-2|$ defines a topological
diffuseness index for diffuse plate boundaries. For the present-day PB2002 network the
hypotheses are measured to be non-vacuous, with $\delta_0(0.15)\approx2.8$~km. The
theorem is the mathematical foundation of the topological audit of plate
reconstructions presented in a companion paper (Kim, 2026).
\end{abstract}

\medskip\noindent\textbf{Relation to the companion paper.} The companion paper
(Kim, 2026; submitted to \textit{Geoscience Frontiers}) applies Euler's polyhedral
formula $V-E+F=2$ to the plate boundary network as a global consistency test for plate
reconstructions, audits published models against it, and uses the theorem proved here
to extend the test from sharp to diffuse boundaries through the topological
diffuseness index. That paper states the theorem, its hypotheses and its numerical
demonstration; this paper contains the proofs. Section~1 fixes notation and the
four-layer structure of the argument; Sections~2--4 establish pointwise convergence,
the CW structure and the classical Euler identity; Section~5 proves the convergence of
the weighted Euler characteristic on the offset cover; Section~6 transfers it to the
normalized field and defines the diffuseness index; Section~7 states the master
theorem; Section~8 records the measured constants for the present-day Earth; and
Section~9 relates the construction to persistent homology.

\tableofcontents
\bigskip

\section{Overview and notation}

\subsection{Preface}

The theory of plate tectonics operates in two registers that have never been formally connected. In the \textit{discrete} register, plates are rigid polygons on a sphere, plate boundaries are their shared edges, and triple junctions are the points where three boundaries meet. This discrete geometry is governed, as we show, by Euler's polyhedral formula $V - E + F = 2$. In the \textit{continuous} register, Bercovici \& Wessel (1994) replaced discontinuous plate boundaries with smooth shape functions $S_i^{\delta^*}$ that interpolate continuously between 1 (inside plate $i$) and 0 (outside plate $i$), parameterized by a boundary half-width $\delta^*$.

These two pictures describe the same physical system --- Earth's lithosphere --- yet no mathematical bridge has connected them. \textbf{The central claim of this paper is that the Bercovici--Wessel continuous model converges, as $\delta^* \to 0$, to the discrete plate boundary network, and that this convergence carries the Euler characteristic $\chi = 2$ along with it.} In other words, $V - E + F = 2$ is not merely a combinatorial curiosity: it is the discrete skeleton to which every continuous plate model collapses in the sharp-boundary limit.

\subsection{Notation and setup}

Let $S^2$ denote the unit 2-sphere (representing the surface of the Earth after radial projection). Throughout this paper:

\begin{center}
\begin{tabular}{l p{9.5cm}}
\toprule
Symbol & Meaning \\
\midrule
$N$ & Total number of tectonic plates \\
$P_i$ & Closed region on $S^2$ occupied by plate $i$, $i = 1, \ldots, N$ \\
$\text{int}(P_i)$ & Interior of plate $i$ \\
$\partial P_i$ & Boundary of plate $i$ (plate boundary arcs) \\
$\delta^*$ & Boundary half-width parameter (a length: radians of pseudo-angle in the original B\&W construction; geodesic distance, quoted in km, for the field of eq.~3) \\
$\Delta^i(p)$ & Pseudo-distance from point $p \in S^2$ to the interior reference point of plate $i$ (original B\&W construction; superseded by $d_i$, eq. 3) \\
$\Delta_b^i(\lambda)$ & Plate-boundary function: pseudo-distance from the interior reference to the boundary, as a function of the pseudo-angle $\lambda$ (original B\&W construction; superseded by $d_i$, eq. 3) \\
$d_i(p)$ & Signed geodesic (great-circle) distance from $p$ to $\partial P_i$: positive inside $P_i$, negative outside (eq. 3) \\
$\widetilde S_i^{\delta^*}$ & Unnormalized generalized shape function built from $d_i$ (eq. 3); $S_i^{\delta^*}$ is its partition-of-unity normalization \\
$c(\cdot)$ & Number of connected components of a subset of $S^2$, with $c(\emptyset) = 0$ (eq. 7) \\
$\rho$ & Geometric separation radius (Lemma 1): no four plates within $\rho$ of a point; distinct co-bounding arcs and distinct junctions $\ge \rho$ apart \\
$\delta_0(\tau)$ & Threshold half-width below which $\chi_w^{\delta^*}(\tau) = 2$ exactly (Theorem 2) \\
$F$ & Number of 2-cells (faces); $F=F_{\text{cell}}=\sum_i c(\operatorname{int}P_i)$, equal to the plate count $N$ under Hypothesis~H2 \\
$E$ & Number of plate-boundary segments (edges in the graph) \\
$V$ & Number of triple junctions (vertices in the graph) \\
$\chi$ & Euler characteristic of the graph $= V - E + F$ \\
$H_i(p)$ & Indicator function of plate $i$: $= 1$ if $p \in \text{int}(P_i)$, $= 0$ otherwise \\
$\tau$ & A threshold parameter, $\tau \in (0, 1/3)$ \\
\bottomrule
\end{tabular}
\end{center}

\subsection{The four layers of the proof: an overview}

The argument is organized into four layers, each building on the previous. Before presenting the details, we explain \textbf{why each layer is necessary} --- what specific logical gap it closes.

\begin{enumerate}
  \item \textbf{Layer 1: Pointwise Convergence.} ``The B\&W functions converge to indicator functions as $\delta^* \to 0$.''
    \begin{description}
      \item[{\normalfont\scshape Why needed:}] Without this, the continuous and discrete models are unrelated objects. This is the foundational bridge.
    \end{description}
  \item \textbf{Layer 2: CW Complex Structure.} ``The discrete plate system forms a valid CW decomposition of $S^2$.''
    \begin{description}
      \item[{\normalfont\scshape Why needed:}] Euler's formula applies to CW complexes, not to arbitrary point sets. We must earn the right to use it.
    \end{description}
  \item \textbf{Layer 3: Euler's Formula (Main Result).} ``$V - E + F = \chi(S^2) = 2$ for any CW decomposition of $S^2$.''
    \begin{description}
      \item[{\normalfont\scshape Why needed:}] This is the core topological fact. It is classical (Hatcher 2002, \S2.2); the Layer~3 section below states it with citation and records the trivalent corollaries, rather than reproving it.
    \end{description}
  \item \textbf{Layer 4: Convergence of the Weighted Euler Characteristic.} ``$\chi_w^{\delta^*}(\tau) \to 2$ as $\delta^* \to 0$.''
    \begin{description}
      \item[{\normalfont\scshape Why needed:}] This connects the continuous B\&W quantity to the discrete Euler formula, enabling the TDI definition and completing the bridge.
    \end{description}
\end{enumerate}

\section{Layer 1: pointwise convergence of the shape functions}

\subsection{Why this layer is necessary}

Bercovici \& Wessel (1994) introduced their shape functions $S_i^{\delta^*}$ as a computational device to avoid the singularities that arise when plate boundaries are treated as discontinuous. Their paper is concerned with kinetic-energy partitioning, not with topology. Consequently, they never asked: \textit{what happens to these functions when the boundary width goes to zero?}

This question is logically prior to everything else in our proof. If the answer were ``the functions do not converge to anything useful,'' there would be no bridge between the continuous and discrete models. \textbf{Layer 1 establishes that the answer is exactly right:} the functions converge, almost everywhere, to the indicator functions of the discrete plates. This is the moment when the continuous model ``remembers'' that it was always an approximation to a discrete reality.

The phrase \textit{almost everywhere} (a.e.) is important. Convergence fails on the measure-zero set of plate boundaries $\partial P_i$, precisely because those boundaries are where the transition from $0$ to $1$ occurs. This is not a defect --- it is correct behavior, because a boundary point genuinely lies between two plates.

\subsection{Setup: the Bercovici--Wessel shape function}

Following Bercovici \& Wessel (1994, eq.\ 25), the shape function for plate $i$ is:

\begin{equation}
S_i^{\delta^*}(\theta, \phi) = \frac{1}{2}\!\left[1 + \tanh\!\left(\frac{\Delta_b^i(\lambda) - \Delta^i(\theta, \phi)}{\delta^*}\right)\right] \tag{1}
\end{equation}

where:
\begin{itemize}
  \item $(\theta, \phi)$ are colatitude and longitude (the position of point $p$),
  \item $\lambda = \lambda(p)$ is the pseudo-angle from the interior reference point $(\theta_0', \phi_0')$ to $p$,
  \item $\Delta^i(p) = \sqrt{(\phi' - \phi_0')^2 + (\theta' - \theta_0')^2}$ is the pseudo-distance (treated as Cartesian in the rotated frame),
  \item $\Delta_b^i(\lambda)$ is the plate-boundary function, giving the pseudo-distance to the boundary in the direction $\lambda$.
\end{itemize}

\medskip\noindent\textbf{Interpretation of the argument of $\tanh$:} The quantity

\[
A_i(p) := \Delta_b^i(\lambda(p)) - \Delta^i(p)
\]

is \textbf{positive} when $p$ is inside plate $i$ (the boundary is farther away than the point itself, in the pseudo-radial direction) and \textbf{negative} when $p$ is outside plate $i$. At the boundary, $A_i(p) = 0$ and $S_i^{\delta^*} = 1/2$.

The full partition-of-unity normalization is:

\begin{equation}
\sum_{i=1}^{N} S_i^{\delta^*}(p) = 1 \quad \text{for all } p \in S^2 \tag{2}
\end{equation}

This is imposed by normalization, or holds approximately in the original B\&W model; we require it exactly here.

\subsection{Generalized shape function: signed geodesic distance}

The B\&W construction (1) presupposes a pseudo-radial parameterization of each plate from an interior reference point. We generalize it by replacing the pseudo-radial coordinate with the \textbf{signed geodesic distance to the plate boundary}. Define, for each plate $i$,

\[
d_i(p) := \begin{cases} +\,\dist_{S^2}(p, \partial P_i) & p \in P_i \\ -\,\dist_{S^2}(p, \partial P_i) & p \notin P_i \end{cases}
\]

where $\dist_{S^2}$ is geodesic (great-circle) distance. Three elementary properties (each immediate from the definition and the compactness of $\partial P_i$):

\begin{enumerate}
  \item $d_i$ is continuous on $S^2$ (indeed 1-Lipschitz on each branch, and both branches vanish on $\partial P_i$);
  \item $d_i(p) = 0$ if and only if $p \in \partial P_i$ (the interior of $P_i$ and the exterior are open);
  \item $\sign(d_i) = +1$ on $\text{int}(P_i)$ and $-1$ on $S^2 \setminus P_i$ --- the sign carries exactly the indicator parity that drives Layer 1.
\end{enumerate}

The \textbf{unnormalized} generalized shape function is

\begin{equation}
\widetilde S_i^{\delta^*}(p) := \frac{1}{2}\!\left[1 + \tanh\!\left(\frac{d_i(p)}{\delta^*}\right)\right] \tag{3}
\end{equation}

and the field used in everything that follows is its normalization to the partition of unity (2):

\[
S_i^{\delta^*}(p) := \frac{\widetilde S_i^{\delta^*}(p)}{\sum_{j=1}^{N} \widetilde S_j^{\delta^*}(p)}
\]

The normalization is well defined: every $p \in S^2$ lies in at least one closed plate $P_m$, so $d_m(p) \geq 0$ and $\widetilde S_m^{\delta^*}(p) \geq 1/2$; hence the denominator is bounded below by $1/2$ everywhere (and the normalized field is continuous, so the level sets (5) remain open). With this normalization, eq.\ (2) holds \textit{exactly by construction}, discharging the caveat ``imposed by normalization, or holds approximately in the original B\&W model'' attached to eq.\ (2).

Eq.\ (3) keeps the two features of (1) that the proof actually uses --- the $\tanh$ profile with half-width $\delta^*$, and an argument whose sign distinguishes inside from outside --- while removing the star-shapedness assumption hidden in the pseudo-radial coordinate (see the Remark at the end of this layer).

\subsection{Proposition 1 (pointwise convergence, generalized field)}

\medskip\noindent\textbf{Statement.} For the field (3) (normalized as above), the conclusion (4) holds verbatim: for each plate $i$ and each $p \in S^2 \setminus \bigcup_i \partial P_i$,

\[
\lim_{\delta^* \to 0} S_i^{\delta^*}(p) = H_i(p), \qquad \text{i.e.} \qquad S_i^{\delta^*} \xrightarrow{\text{a.e.}} H_i \quad (\delta^* \to 0).
\]

where, as in the discrete model,

\begin{equation}
H_i(p) := \begin{cases} 1 & \text{if } p \in \text{int}(P_i) \\ 0 & \text{if } p \notin P_i \end{cases} \tag{4}
\end{equation}

\begin{proof}

The original Bercovici--Wessel argument used exactly two properties of the $\tanh$ argument $A_i(p) = \Delta_b^i(\lambda(p)) - \Delta^i(p)$: that it is positive (some $c > 0$) for $p \in \text{int}(P_i)$, and negative ($-c < 0$) for $p \notin P_i$. By properties 2--3 above, $d_i(p)$ has the same sign structure, with $c = |d_i(p)| > 0$ off the boundary set. Cases 1--3 of that argument therefore go through verbatim for $\widetilde S_i^{\delta^*}$ with $A_i$ replaced by $d_i$:

\[
\left|\widetilde S_i^{\delta^*}(p) - H_i(p)\right| \leq e^{-2|d_i(p)|/\delta^*} \quad \text{for } p \notin \partial P_i,
\]

the same exponential rate. For the normalized field: fix $p$ off the boundary network and let $D(p) := \dist_{S^2}(p, \bigcup_j \partial P_j) > 0$. Then $|d_j(p)| \geq D(p)$ for every $j$, so each $\widetilde S_j^{\delta^*}(p)$ is within $e^{-2D(p)/\delta^*}$ of $H_j(p)$, the denominator is within $N e^{-2D(p)/\delta^*}$ of $\sum_j H_j(p) = 1$, and (using denominator $\geq 1/2$)

\[
\left|S_i^{\delta^*}(p) - H_i(p)\right| \leq 2(N+1)\, e^{-2D(p)/\delta^*}.
\]

So the normalized field converges a.e.\ to $H_i$ with the same exponential rate in $1/\delta^*$, up to a fixed constant. On the boundary set (measure zero) convergence fails exactly as in Case 3 of that argument: at an interior point of an arc shared by plates $i, j$, $S_i^{\delta^*} \to 1/2$; at a trivalent junction, $S_i^{\delta^*} \to 1/3$ (see Lemma 2 below for the uniform version of these statements).
\end{proof}

\medskip\noindent\textbf{Corollary 1 (partition of unity in the limit).} With the normalization above, eq.\ (2) is exact at every $\delta^*$, so the limit statement $\sum_i H_i = 1$ a.e.\ follows as before (dominated convergence, $0 \leq S_i^{\delta^*} \leq 1$).

\subsection{Remark (why the generalization is needed)} %

For a plate that is \textit{star-shaped} with respect to its interior reference point, the pseudo-radial argument $A_i(p) = \Delta_b^i(\lambda(p)) - \Delta^i(p)$ of eq.\ (1) is a single-valued signed coordinate vanishing exactly on $\partial P_i$ with the same sign structure as $d_i$, and near a boundary point where the radial ray meets $\partial P_i$ transversally, $A_i(p) = d_i(p)/\cos\gamma + O(d_i^2)$, where $\gamma$ is the angle between the ray and the inward boundary normal. Thus for star-shaped plates eq.\ (1) coincides with eq.\ (3) to first order near the boundary, up to a direction-dependent effective half-width $\delta^*_{\mathrm{eff}} = \delta^* \cos\gamma$; all limit statements of this document are insensitive to such a bounded reparameterization of $\delta^*$.

For plates that are \textit{not} star-shaped about any practical reference point --- the geologically common case (concave plate outlines: Pacific, Australia, Eurasia, \dots) --- the boundary function $\Delta_b^i(\lambda)$ is not single-valued: a pseudo-radial ray may cross $\partial P_i$ several times, and points geodesically close to the boundary but angularly behind a concavity acquire $A_i$ values belonging to the wrong boundary crossing. The level sets of (1) then misrepresent the boundary geometry in the shadow of each concave wedge (``wedge shadowing''). Empirically, in the PB2002 implementation of eq.\ (1), wedge shadowing corrupted the field at \textbf{52 of the 100 triple junctions}, destroying the $\min \to 1/3$ witness on which Step 2 of Theorem 2 relies. Eq.\ (3) eliminates the artifact at its source: the geodesic distance to a compact boundary is intrinsically single-valued and reference-point-free. This is the motivation for adopting (3); no result of Layers 2--3 is affected, and Layer 1 carries over by Proposition 1.

\section{Layer 2: CW-complex structure of the plate system}

\subsection{Why this layer is necessary}

Knowing that $S_i^{\delta^*} \to H_i$ is a statement about \emph{functions}. To invoke Euler's formula, we need a statement about \emph{topology}. Specifically, Euler's formula $V - E + F = 2$ is a theorem about \textbf{CW complexes} --- a rigorous topological concept that generalizes the notion of polyhedra. A CW complex is a space built by attaching \emph{cells} of increasing dimension: 0-cells (points), 1-cells (arcs), 2-cells (disks).

\textbf{Layer 2 is necessary because we must prove that the plate system is a CW complex of the right kind before Euler's formula can be applied.} This is not automatic. An arbitrary collection of regions on $S^2$ need not form a CW complex: boundaries could be fractal, cells could fail to be homeomorphic to open disks, or the gluing maps could be pathological. We must verify all the required conditions.

In our context, the key structural fact comes from geophysics: \textbf{all junctions of plate boundaries are triple junctions} (three boundaries meeting at a single point). This is not a mathematical assumption but a dynamical consequence established by McKenzie \& Morgan (1969): quadruple junctions are instantaneously unstable and immediately split into two triple junctions. This physical fact is what gives our graph its regular degree-3 vertex structure, which in turn tightly constrains the relationship between $V$, $E$, and $F$.

\subsection{Standing structural hypothesis (H2)}

Everything in Layers 2--4 rests on one \textbf{explicit structural hypothesis} about the sharp plate mosaic $\{P_i\}_{i=1}^{N}$. We state it as a hypothesis, \textbf{not} as something to be derived, for a precise reason given below (a connected arc-bounded region need not be a disk). For a digital plate model the hypothesis holds by construction or is directly checkable from the polygon data; where a real reconstruction violates it, the reconstruction audit of the companion paper (Kim, 2026) is the instrument that detects the violation --- so making H2 explicit is what turns the framework into a \emph{test}.

\medskip\noindent\textbf{Hypothesis H2 (finite regular trivalent disk-CW decomposition of $S^2$).} The sharp plate mosaic forms a finite regular cell (CW) decomposition $\mathcal{K}$ of $S^2$ in which:

\begin{enumerate}
\item each plate region $P_i$ is homeomorphic to a closed disk $\overline{B^2}$;
\item each interior $\text{int}(P_i)$ is an open disk $B^2$;
\item for $i\ne j$, $P_i\cap P_j$ is either empty or a finite union of boundary arcs and junction points;
\item every open boundary arc is incident to \textbf{exactly two distinct} plates;
\item every junction is incident to \textbf{exactly three distinct} plates (trivalent);
\item no boundary arc has a non-junction endpoint (every arc closes at junctions);
\item the union of all cells is $S^2$ (the $P_i$ cover $S^2$ with pairwise-disjoint interiors).
\end{enumerate}

Under H2 the discrete cell counts $(V,E,F)$ are well defined and the discrete Euler result of Layer 3 applies verbatim; the trivalent clause (5), via the handshaking argument, then yields the corollaries $E=3(F-2)$, $V=2(F-2)$ and the discrete $\chi=V-E+F=2$.

\medskip\noindent\textbf{Why this is a hypothesis and not a theorem.} Clauses (1)--(2) are \emph{not} delivered by the Jordan curve theorem. Jordan separation on $S^2$ states only that a simple closed curve $\partial P_i$ separates the sphere into two components; it says nothing about a region whose boundary is a \emph{union of arcs} meeting at junctions. Such a connected arc-bounded region can be an \textbf{annulus} --- for instance when one plate \textbf{encloses} another (a microplate) --- so its boundary network has two components and the region is not a disk. Disk-likeness must therefore be \textbf{assumed} (clauses 1--2) and, on real data, \textbf{checked}.

\medskip\noindent\textbf{Remark (the audit detects where H2 fails --- a demonstrated capability).} The extraction pipeline flags exactly the failure of H2: a boundary 1-skeleton with more than one connected component ($C>1$, satisfying the identity $\chi=1+C$ for a valid embedded graph) signals an isolated boundary loop, i.e.\ an annular face, i.e.\ a region violating clauses (1)--(2). A concrete instance occurs in the audit: in Merdith et al.\ (2021) at \textbf{851 Ma} (persisting 851--853 Ma) the \textbf{West Ethiopian Shield} (PID 77521, area $\approx 0.0100$ sr) is enclosed inside \textbf{Rodinia} (PID 101, area $\approx 5.94$ sr) as a four-arc isolated loop, giving a two-component boundary network with $C=2$ and $\chi=3=1+C$ --- identity-consistent, yet \textbf{not} a disk-CW decomposition. Rodinia is recorded in the model as a single disk-like face but is geometrically an \textbf{annulus} that double-covers the enclosed shield: the surface coverage overcounts by $\approx 0.08\%$ (coverage $=1.0008$), exactly the shield's fractional area, the signature of the enclosed region being covered twice. Thus the disk hypothesis is empirically violated by a published reconstruction, and the audit is the tool that catches it (the companion paper (Kim, 2026), Fig.~5c). H2 is stated openly, not assumed silently, precisely so that the framework can be used to test it on real reconstructions.

\subsection{Background: CW complexes}

A \textbf{CW complex} is a topological space $X$ together with a partition into open cells $\{e_\alpha^n\}$ (indexed by dimension $n$ and label $\alpha$) such that:

\begin{enumerate}
\item Each open $n$-cell $e_\alpha^n$ is homeomorphic to the open $n$-ball $B^n = \{x \in \mathbb{R}^n : |x| < 1\}$.
\item The closure $\overline{e_\alpha^n}$ is the image of a continuous map $\Phi_\alpha : \overline{B^n} \to X$ (the \emph{characteristic map}) whose restriction to the interior is a homeomorphism onto $e_\alpha^n$.
\item The image $\Phi_\alpha(\partial B^n)$ is contained in the union of cells of dimension $< n$ (the \emph{closure-finite} condition).
\item A subset $A \subseteq X$ is closed if and only if $A \cap \overline{e_\alpha^n}$ is closed for each $\alpha$ (the \emph{weak topology} condition).
\end{enumerate}

For a finite CW complex (finite number of cells), conditions 3 and 4 are automatic.

\subsection{Proposition 2 (plate system as CW complex)}

\noindent\textbf{Statement.} In the sharp-boundary limit $\delta^* \to 0$, the plate system $\{P_i\}$ induces a finite CW decomposition of $S^2$, denoted $\mathcal{K}$, with:

\begin{itemize}
\item \textbf{0-cells}: the set of triple junctions $\mathcal{V} = \{v_1, \ldots, v_V\}$, each of degree 3,
\item \textbf{1-cells}: the open boundary arcs $\mathcal{E} = \{e_1, \ldots, e_E\}$, each connecting two (not necessarily distinct) triple junctions,
\item \textbf{2-cells}: the open plate interiors $\mathcal{F} = \{\text{int}(P_1), \ldots, \text{int}(P_F)\}$.
\end{itemize}

\begin{proof}
We verify each CW condition.

\medskip\noindent\textbf{Step 1: Identification of cells.}
From Layer 1, the limit indicator functions $H_i$ partition $S^2$ (a.e.) into:
\begin{itemize}
\item Open regions $\text{int}(P_i)$: these will be the 2-cells.
\item Boundary curves $\partial P_i \cap \partial P_j$ (for $i \neq j$): these will provide the 1-cells.
\item Points $\partial P_i \cap \partial P_j \cap \partial P_k$ (for distinct $i,j,k$): these will be the 0-cells.
\end{itemize}

\medskip\noindent\textbf{Step 2: Degree-3 condition at vertices.}
The McKenzie--Morgan theorem (1969) establishes that, under plate-tectonic kinematics, the number of boundaries meeting at any junction is generically 3. This is because the kinematic stability conditions for a junction of degree $d$ reduce to $d - 3$ constraint equations on the velocity field. For $d \geq 4$, these constraints are generically not satisfied, so degree-$d$ junctions immediately evolve to two degree-3 junctions. Mathematically: every vertex $v \in \mathcal{V}$ has degree $\deg(v) = 3$.

\medskip\noindent\textbf{Step 3: Verification of the CW conditions for each cell type.}

\emph{For 0-cells}: each triple junction is a point, homeomorphic to $B^0 = \{0\}$. The characteristic map is the inclusion map.

\emph{For 1-cells}: each open boundary arc $e_j$ (a connected component of $\partial P_i \cap \partial P_j$ minus its endpoints) is homeomorphic to the open interval $(0,1) \cong B^1$. Its closure is a closed arc with endpoints at triple junctions, so the boundary $\partial e_j$ is contained in $\mathcal{V}$. The characteristic map $\Phi_j : [0,1] \to S^2$ is a smooth embedding of the arc.

\emph{For 2-cells}: by \textbf{Hypothesis H2} (clauses 1--2, stated in ``Standing structural hypothesis (H2)'' above), each open plate interior $\text{int}(P_i)$ is an open disk $B^2$, and the characteristic map $\Phi_i : \overline{B^2} \to P_i$ extends the homeomorphism over the boundary arcs. We emphasize that disk-likeness is \textbf{assumed} (H2), \textbf{not} derived from the Jordan curve theorem: Jordan separation gives only that $\partial P_i$ separates $S^2$, and a connected arc-bounded region can be an \textbf{annulus} when one plate encloses another (a microplate). The audit of the companion paper (Kim, 2026) detects precisely such violations as a disconnected boundary 1-skeleton ($C>1$); see the Remark in ``Standing structural hypothesis (H2)'' for the Merdith et al.\ (2021) 851 Ma instance (Rodinia recorded as a disk but geometrically an annulus enclosing the West Ethiopian Shield). Under H2 the CW condition for 2-cells holds.

\emph{Closure-finite condition}: The closure of each 2-cell $P_i$ is $P_i = \text{int}(P_i) \cup \partial P_i$. The boundary $\partial P_i$ consists of finitely many 1-cells and 0-cells (since the plate system is finite).

\emph{Weak topology}: The number of cells is finite, so the weak topology equals the subspace topology from $S^2$.

\medskip\noindent\textbf{Step 4: Completeness of the partition.}
We must verify that $S^2 = \mathcal{V} \cup \bigsqcup_j e_j \cup \bigsqcup_i \text{int}(P_i)$ (disjoint union). By the partition-of-unity limit (Corollary 1), every point of $S^2$ lies in the interior of exactly one plate (a.e.), or on exactly one boundary arc (if it is a non-junction boundary point), or at exactly one triple junction. This gives a partition of all of $S^2$ into the three types of cells. \checkmark

\medskip\noindent\textbf{Faces are 2-cells, not plate labels.} Throughout, $F$ denotes the number of \textbf{2-cells}, i.e.
\[
F \;=\; F_{\text{cell}} \;:=\; \sum_{i=1}^{N} c\!\left(\mathrm{int}\,P_i\right),
\]
the total number of connected components of the plate interiors (here $N$ is the number of plate labels and $c(\cdot)$ counts connected components, as in eq.~7). Under the standing \textbf{Hypothesis H2} each plate $P_i$ is a single disk, so every label is connected and $F_{\text{cell}} = N =$ the number of plates; this is the idealized case in which the reading ``$F$ = number of plates'' is exact. When H2's per-label connectedness is \textbf{relaxed} --- as a digital reconstruction may assign one plate label to geographically disconnected fragments --- the Euler-valid face count is the 2-cell count $F_{\text{cell}} = \sum_i c(\mathrm{int}\,P_i)$, and the identity reads $V - E + F_{\text{cell}} = 2$. The proof machinery already counts componentwise: the component-counted weighted characteristic (7) sums $c(U_i)$ over labels, and Step 1 (Layer 4) shows $c(U_i) = c(\mathrm{int}\,P_i)$ in the sharp regime, so it recovers $F_{\text{cell}}$ automatically. ($F_{\text{cell}}$ is the robust count; H2 is the idealized special case $F_{\text{cell}} = N$.)

\medskip\noindent\textbf{Conclusion.} The triple $(\mathcal{V}, \mathcal{E}, \mathcal{F})$ defines a finite CW decomposition of $S^2$.
\end{proof}

\section{Layer 3: Euler's formula $V-E+F=2$}

Having established in Layer~2 that the plate system is a CW complex, this layer
supplies the core topological fact, which we treat as standard and cite rather
than prove.

\medskip\noindent\textbf{Theorem 1 (Euler's formula).} For any finite CW
decomposition of $S^2$,
\[
  V-E+F=2
\]
(Hatcher 2002, \textit{Algebraic Topology}, \S2.2).

\medskip\noindent\textbf{Euler--Poincar\'e computation.}
The Euler characteristic is a homological invariant. For the $2$-sphere,
\[
  \chi(S^2)=\operatorname{rank}H_0-\operatorname{rank}H_1+\operatorname{rank}H_2
          =1-0+1=2,
\]
and since $\chi$ equals the alternating cell count $c_0-c_1+c_2=V-E+F$ of any CW
decomposition, we obtain $V-E+F=2$.

\medskip\noindent\textbf{Trivalent (handshaking) corollaries.}
By the handshaking identity $\sum_{v}\deg(v)=2E$, and because every vertex of the
plate system has degree $3$ (Layer~2), we get $3V=2E$. Combining with
$V-E+F=2$ yields the constraints underlying the missing-plate bound:
\[
  E=3(F-2),\qquad V=2(F-2).
\]

\medskip\noindent\textbf{Face count and the degenerate case.}
The face count is the $2$-cell count,
$F=F_{\text{cell}}=\sum_i c(\operatorname{int}P_i)$, where $c(\cdot)$ counts
connected components; Hypothesis~H2 is the idealized connected case in which each
plate interior is a single open disk, so $F_{\text{cell}}=N$ (the number of
plates). The two-plate configuration is a valid CW decomposition of $S^2$
($V=1$, $E=1$, $F=2$, $\chi=2$), so it is \emph{not} excluded by the Euler
characteristic. It is excluded from the all-trivalent model by trivalency
together with the minimum-face-degree relation $3F\le 2E$, which with $E=3(F-2)$
gives $3F\le 6(F-2)$, i.e. $F\ge 4$.

\medskip\noindent\textbf{Gauss--Bonnet motivation.}
Geometrically (Gauss--Bonnet; Gauss 1827), $\chi$ is the total curvature divided by $2\pi$: for the unit
sphere $\frac{1}{2\pi}\int_{S^2}K\,dA=2$, which motivates the weighted Euler
characteristic of Layer~4.

\medskip
The three full independent proofs (homological, handshaking, and Gauss--Bonnet)
are standard and are omitted here.

\section{Layer 4: convergence of the weighted Euler characteristic}

\subsection{Why this layer is necessary}

The first three layers established:
\begin{itemize}
  \item (Layer 1) The B\&W functions $S_i^{\delta^*}$ converge pointwise to indicator functions.
  \item (Layer 2) The limit plate system is a CW complex.
  \item (Layer 3) Any CW decomposition of $S^2$ satisfies $V - E + F = 2$.
\end{itemize}

But these three layers, taken together, only give us the following: \textit{in the limit $\delta^* \to 0$, the plate system satisfies Euler's formula.} This is a statement about the limit, not about the continuous model at finite $\delta^*$.

\medskip\noindent\textbf{Layer 4 is necessary for two reasons:}

\begin{enumerate}
  \item \textbf{For real Earth:} The Earth has finite boundary widths --- $\delta^* \neq 0$ in practice. We need a quantity that is defined at finite $\delta^*$, reduces to $\chi = 2$ in the limit, and deviates from 2 in a physically meaningful way for diffuse boundaries. This is the \textbf{Weighted Euler Characteristic} $\chi_w^{\delta^*}(\tau)$.

  \item \textbf{For mathematical completeness:} We need to show that the \textit{convergence} $S_i^{\delta^*} \to H_i$ implies the convergence $\chi_w^{\delta^*}(\tau) \to V - E + F = 2$. This requires a careful analysis of how the overlap structure of the shape functions changes as $\delta^* \to 0$, which is the content of the \v{C}ech complex and nerve theorem machinery.
\end{enumerate}

The \textbf{Topological Diffuseness Index} $\text{TDI}(\delta^*, \tau) = |\chi_w^{\delta^*}(\tau) - 2|$ then measures how far a given plate system is from being topologically sharp --- a geophysically novel quantity with no prior analog in the literature.

\subsection{Definition: level sets and the \v{C}ech complex}

For a fixed threshold $\tau \in (0, 1/3)$ and boundary half-width $\delta^*$, define the \textbf{level-set open cover} of $S^2$:

\begin{equation}
\mathcal{U}^{\delta^*,\tau} = \left\{U_i^{\delta^*,\tau}\right\}_{i=1}^N, \quad U_i^{\delta^*,\tau} := \left\{p \in S^2 : S_i^{\delta^*}(p) > \tau\right\} \tag{5}
\end{equation}

Each $U_i^{\delta^*,\tau}$ is an open subset of $S^2$ (since $S_i^{\delta^*}$ is continuous): it is the region where plate $i$ has membership greater than $\tau$.

\subsection{The unnormalized offset cover (the topology-counting cover)}

The normalized level sets (5), $U_i^{\delta^*,\tau}=\{p:S_i^{\delta^*}(p)>\tau\}$, are the natural objects for the \textbf{physical} fuzzy-membership reading: each $S_i^{\delta^*}$ is a partition-of-unity membership (eq. 2), and the family $\{U_i^{\delta^*,\tau}\}$ remains bounded --- overlap counts stay finite and the TDI stays well-behaved --- even as $\delta^*$ grows into the diffuse regime. For the \textbf{topology count} we use instead the \textbf{unnormalized offset cover}, on which the convergence argument is standard differential topology.

Because $\widetilde{S}_i^{\delta^*}(p)=\tfrac12\!\left[1+\tanh\!\left(d_i(p)/\delta^*\right)\right]$ is a strictly increasing function of the signed geodesic distance $d_i$, its level set is \textbf{exactly} a one-sided signed-distance offset of the plate:

\begin{equation}
\widetilde{U}_i^{\delta^*,\tau} \;:=\; \left\{p\in S^2 : \widetilde{S}_i^{\delta^*}(p) > \tau\right\} \;=\; \left\{p\in S^2 : d_i(p) > -\,r_\tau\right\}, \qquad r_\tau \;:=\; \delta^*\,\artanh(1-2\tau). \tag{6}
\end{equation}

Equivalently $\widetilde{U}_i^{\delta^*,\tau}=P_i \cup \{p: \dist_{S^2}(p,P_i)<r_\tau\}$, the open $r_\tau$-neighborhood (one-sided offset) of the closed plate $P_i$, with $r_\tau\to 0$ as $\delta^*\to 0$. The offset radius $r_\tau$ is defined and positive for every $\tau\in(0,\tfrac12)$ --- a \textbf{wider} band than the $\tau\in(0,1/3)$ used for the witness bound (W1); the whole reported ladder $\tau\in\{0.15,0.25,0.30\}$ lies inside it, with

\[
\frac{r_\tau}{\delta^*}=\artanh(1-2\tau)\;\approx\; 0.867 \,/\, 0.549 \,/\, 0.424 \qquad (\tau=0.15\,/\,0.25\,/\,0.30),
\]

so at $\delta^*=50$ km, $r_\tau\approx 43.4 \,/\, 27.5 \,/\, 21.2$ km --- well inside the documented $\sim 50$ km sharp band. We write $\widetilde{\chi}_w^{\delta^*}(\tau)$ for the weighted Euler characteristic (7) computed on the offset cover $\{\widetilde{U}_i^{\delta^*,\tau}\}$ in place of $\{U_i^{\delta^*,\tau}\}$.

\medskip\noindent\textbf{Two-track convention (used throughout Layer 4 from here on).} The sharp-limit \textit{topology} is proved on the \textbf{offset cover} $\{\widetilde{U}_i\}$, where the connectedness lemmas reduce to positive-reach / tubular-neighborhood theory (Theorem 2, Lemma 3 below). The \textbf{TDI numerics} and the physical B\&W interpretation are reported on the \textbf{normalized cover} $\{U_i\}$, which is the field that stays bounded as $\delta^*$ grows (the offset cover \textit{diverges} in the diffuse limit --- see the Remark after Corollary 2). The Comparison Lemma (Lemma 4) shows the two covers have the \textbf{same component incidences} for $\delta^*<\delta_0(\tau)$, so $\widetilde{\chi}_w^{\delta^*}(\tau)=\chi_w^{\delta^*}(\tau)$ in that range and the sharp-limit result transfers to the normalized $\chi_w$ used in the TDI. No numerical result of the companion paper changes.

The \textbf{\v{C}ech complex} $\check{C}(\mathcal{U}^{\delta^*,\tau})$ of this cover is the abstract simplicial complex with:
\begin{itemize}
  \item A 0-simplex $[i]$ for each $U_i^{\delta^*,\tau}$,
  \item A 1-simplex $[i,j]$ whenever $U_i^{\delta^*,\tau} \cap U_j^{\delta^*,\tau} \neq \emptyset$,
  \item A 2-simplex $[i,j,k]$ whenever $U_i^{\delta^*,\tau} \cap U_j^{\delta^*,\tau} \cap U_k^{\delta^*,\tau} \neq \emptyset$.
\end{itemize}

(Higher-dimensional simplices vanish because at most 3 level sets overlap significantly, as we prove below.)

\subsection{Definition: weighted Euler characteristic (component counting)}

For a (possibly empty) subset $A \subseteq S^2$, let $c(A)$ denote the number of connected components of $A$, with $c(\emptyset) = 0$. The \textbf{weighted Euler characteristic} of the cover (5) is

\begin{equation}
\chi_w^{\delta^*}(\tau) := \sum_{i} c\!\left(U_i^{\delta^*,\tau}\right) \;-\; \sum_{i<j} c\!\left(U_i^{\delta^*,\tau} \cap U_j^{\delta^*,\tau}\right) \;+\; \sum_{i<j<k} c\!\left(U_i^{\delta^*,\tau} \cap U_j^{\delta^*,\tau} \cap U_k^{\delta^*,\tau}\right) \tag{7}
\end{equation}

Eq. (7) refines the simple \v{C}ech count: it replaces each indicator ``$\neq \emptyset$ contributes 1'' with the number of connected components of the same region. When every $U_i$ is nonempty and connected, every nonempty pairwise intersection is connected, and every nonempty triple intersection is connected, (7) reduces to the simple count; the two definitions differ exactly when intersections are disconnected, which is the generic situation for real plate networks (see the Remark on hypothesis (G) below).

\medskip\noindent\textbf{Which cover carries the topology count.} Per the two-track convention (eq. 6 above), the \textit{sharp-limit topology} count $\chi_w$ --- i.e. the equality with $V-E+F$ in Theorem 2 --- is established on the \textbf{unnormalized offset cover} $\{\widetilde{U}_i^{\delta^*,\tau}\}$, written $\widetilde{\chi}_w^{\delta^*}(\tau)$; the \textbf{normalized cover} $\{U_i^{\delta^*,\tau}\}$ of (5) carries the \textit{TDI numerics} and the physical B\&W reading. The two agree, $\widetilde{\chi}_w^{\delta^*}(\tau)=\chi_w^{\delta^*}(\tau)$, for $\delta^*<\delta_0(\tau)$ by the Comparison Lemma (Lemma 4), so the component-counted definition (7) is the right invariant on either cover in the sharp regime.

\medskip\noindent\textit{The component count is not a simplicial Euler characteristic.} The component-counted $\chi_w$ of (7) is \textbf{not} the Euler characteristic of the simplicial \v{C}ech complex: a disconnected pairwise intersection contributes several units to (7) but only one 1-simplex to the simplicial nerve. The homotopy-level object computed by (7) is the \textit{multinerve} of the cover; see the Remark on the nerve interpretation below. A nonemptiness-only (max--min) formulation detects only nonemptiness; its component-aware replacement is

\begin{equation}
c\!\left(U_i \cap U_j\right) = c\!\left(\left\{p \in S^2 : \min\!\left(S_i^{\delta^*}(p), S_j^{\delta^*}(p)\right) > \tau\right\}\right), \tag{8}
\end{equation}

which is an identity of sets ($U_i \cap U_j = \{\min(S_i, S_j) > \tau\}$), with the analogous identity for triples (the minimum taken over three fields). Computationally, the counts in (7) are obtained by connected-component labeling of these regions on the discretization grid, with longitude wrap-around and polar rows handled on the sphere.

\subsection{Remark (hypothesis (G) and why the simple count fails on real networks)}

The simple-count form of Theorem 2 (Steps 1--2) shows that the simple pairwise count converges to the number of \textit{adjacent plate pairs} and the simple triple count to the number of \textit{plate triples meeting at some junction}. Its conclusion ``$= E$'' and ``$= V$'' therefore silently assumed:

\medskip\noindent\textbf{Hypothesis (G) (genericity).} Every adjacent pair of plates shares exactly one boundary arc, and every triple of plates meets in at most one junction.

\medskip
Under (G), pairs are in bijection with edges and triples with vertices, and the simple-count statement is correct. \textbf{Real plate networks violate (G).} For PB2002 ($F = 52$, $E = 150$, $V = 100$, all junctions trivalent): there are only \textbf{141} unique adjacent pairs, because \textbf{8} pairs share two or more disjoint boundary arcs (AU--PA shares three), absorbing $150 - 141 = 9$ edges; and only \textbf{94} unique junction triples, because \textbf{6} triples meet at two distinct junctions each, absorbing $100 - 94 = 6$ vertices. The sharp limit of the simple \v{C}ech count on PB2002 is therefore

\[
52 - 141 + 94 = 5 = 2 + 9 - 6 \neq 2,
\]

i.e. the simple \v{C}ech count converges, but to the Euler characteristic of the wrong complex (the simplicial nerve, which collapses parallel edges and repeated triples). Component counting (7) restores the bijection with the cells of the CW decomposition --- each of the 9 extra arcs and 6 extra junctions is recovered as its own component --- and the sharp limit returns to $2$ exactly. (The bookkeeping identity above follows directly from the PB2002 graph counts $F=52$, $E=150$, $V=100$ and the multiplicities recorded above.)

\subsection{Theorem 2 (Convergence of the Weighted Euler Characteristic)}

\medskip\noindent\textbf{Explicit hypotheses.} These refine the Layer-2 standing hypothesis \textbf{H2} (finite regular trivalent disk-CW decomposition of $S^2$): clauses (3)--(7) of H2 supply the combinatorial regularity used below, and the three hypotheses now stated are the \textbf{metric/quantitative strengthening} of H2's topological disk clauses (1)--(2) --- they add the positive-reach, separation, and bounded-sector data needed to control the \textit{offset} cover. They do \textbf{not} replace or contradict H2; (H-reach) is the metric form of ``$P_i$ is a disk,'' (H-separation) refines clause (3), and (H-trivalent-sectors) refines clause (5). Assume, in addition to H2:

\begin{itemize}
  \item \textbf{(H-reach) Positive reach of the smooth boundary, localized to arc interiors, two-sided.} Each plate boundary $\partial P_i$ is piecewise-smooth, smooth on the open interior of each boundary arc, with corners only at the trivalent junctions. On each arc interior $\partial P_i$ has \textbf{two-sided} positive reach in the sense of Federer (1959): there is $r_i>0$ such that every point within geodesic distance $r_i$ of that arc --- \textit{on either side}, i.e. inside $P_i$ and inside its complement --- has a unique nearest point on $\partial P_i$. Set $r_{\min}:=\min_i r_i>0$. (Two-sidedness is used because Step 1 retracts the \textbf{outer} collar of $P_i$ inward; one-sided reach of $P_i$ alone is insufficient.) \textbf{Localization is essential:} a piecewise-geodesic spherical polygon has reach $0$ \textbf{at each corner} (the nearest-point projection is multi-valued in the exterior wedge of any non-smooth corner), so a \textit{global} ``$\mathrm{reach}(P_i)>0$'' is false and is not assumed; the corners (= junctions) are handled separately by (H-trivalent-sectors) and the conical junction model in Lemma 3(3). (H-reach) thus excludes cusps and zero-width slivers on the \textit{smooth part} of $\partial P_i$ only.
  \item \textbf{(H-separation) Geometric separation.} The separation radius $\rho>0$ of \textbf{Lemma 1} holds: no four plates lie within $\rho$ of a point, and any two distinct co-bounding arcs of the same plate pair, and any two distinct junctions, are at geodesic distance $\ge\rho$.
  \item \textbf{(H-trivalent-sectors) Bounded junction sectors (conical corner model).} There is $\theta_0>0$ such that the three boundary sectors at every trivalent junction have opening angle $\ge\theta_0$, and in a fixed ball around each junction (radius $\ge\rho$) the boundary network is $C^{1,1}$-close to the three geodesic rays of its tangent cone; equivalently the incident arcs meet transversally with pairwise angle bounded below (so offset collars cross in a genuine tube, not a pinched neck). This hypothesis is what controls the corners that (H-reach) deliberately excludes; it degrades gracefully as $\theta_0\to 0$ (the sector-width margin $c(\theta_0)\to 0$ and $\delta_0\to 0$) but does not fail while $\theta_0>0$.
\end{itemize}

The fields are (3); the offset cover is (6); $\tau\in(0,\tfrac12)$ is fixed.

\medskip\noindent\textbf{Statement.} There exists $\delta_0=\delta_0(\tau)>0$ such that for all $0<\delta^*<\delta_0$,

\begin{equation}
\widetilde{\chi}_w^{\delta^*}(\tau) \;=\; V-E+F \;=\; 2, \tag{9}
\end{equation}

with $\widetilde{\chi}_w$ the component-counted weighted Euler characteristic (7) evaluated on the offset cover (6). In particular $\lim_{\delta^*\to 0}\widetilde{\chi}_w^{\delta^*}(\tau)=2$; since $\widetilde{\chi}_w$ is integer-valued, the limit is \textbf{attained} below $\delta_0$, not merely approached. By the Comparison Lemma (Lemma 4) the same conclusion holds for the normalized $\chi_w^{\delta^*}(\tau)$ on the (possibly smaller) range $\delta^*<\delta_0(\tau)$, where for the normalized transfer $\delta_0(\tau)$ denotes the smaller, $\tau$-dependent threshold sharpened in Lemma 5 (which adds the junction slot that vanishes as $\tau\uparrow 1/3$).

\medskip\noindent\textbf{Explicit threshold.} The admissible $\delta_0$ is the \textbf{minimum of three competing constraints}, one per hypothesis:
\[
\delta_0(\tau)\;=\;\frac{\min\!\big(r_{\min},\ \rho/2,\ c(\theta_0)\big)}{\artanh(1-2\tau)},\quad\text{equivalently}\quad r_\tau=\delta^*\artanh(1-2\tau)<\min\!\big(r_{\min},\ \rho/2,\ c(\theta_0)\big),
\]
where $r_{\min}$ (H-reach) controls the within-plate retraction, $\rho/2$ (H-separation, \textbf{note the factor $\tfrac12$} --- two $r_\tau$-tubes around cells $\rho$ apart stay disjoint only when $2r_\tau<\rho$) controls between-cell separation, and $c(\theta_0)>0$ (H-trivalent-sectors, a sector-width margin, $c(\theta_0)\propto\rho\sin(\theta_0/2)$ for the worst sector) controls junction connectivity. Because $\artanh(1-2\tau)$ \textbf{decreases} on $(0,\tfrac12)$, $\delta_0$ \textbf{increases} with $\tau$ --- the offset cover is \textit{more} forgiving at larger $\tau$, the exact opposite of the normalized field (which is catastrophic at $\tau=0.30$). This is why the offset proof runs cleanly across the whole ladder.

The theorem is unconditional under the stated hypotheses: the proof rests on the three offset-topology counting statements (L1)--(L3) below, supplied by the Offset Topology Lemma (Lemma 3), together with Lemmas 1 and 2.

\begin{itemize}
  \item \textbf{(L1) Faces.} $c(\widetilde{U}_i)=c(\mathrm{int}\,P_i)$, and (under connected plates) $=1$ per plate; hence $\sum_i c(\widetilde{U}_i)=F$.
  \item \textbf{(L2) Arcs.} $c(\widetilde{U}_i\cap\widetilde{U}_j)=m_{ij}$, the number of distinct arcs shared by plates $i,j$ --- counted correctly even for \textbf{multi-arc pairs} (e.g.\ AU--PA, which share three); hence $\sum_{i<j}c(\widetilde{U}_i\cap\widetilde{U}_j)=E$.
  \item \textbf{(L3) Junctions.} $c(\widetilde{U}_i\cap\widetilde{U}_j\cap\widetilde{U}_k)=m_{ijk}$, the number of junctions at which all three meet --- counting twice-met triples twice; hence $\sum_{i<j<k}c(\widetilde{U}_i\cap\widetilde{U}_j\cap\widetilde{U}_k)=V$.
\end{itemize}

L1--L3 are the three clauses of the \textbf{Offset Topology Lemma (Lemma 3)} below, summed over cells with Lemmas 1 and 2 supplying separation and the witness bound (W1). Lemmas 1 and 2 are proved first; the connectedness statement is the \textbf{Offset Topology Lemma}, proved under (H-reach)/(H-separation)/(H-trivalent-sectors).

\medskip\noindent\textbf{Lemma 1 (geometric separation).} There exists $\rho > 0$ such that: (i) every $p \in S^2$ satisfies $\dist_{S^2}(p, P_j) < \rho$ for at most 3 plates $j$; (ii) any two distinct arcs shared by the same plate pair, and any two distinct junctions, are at geodesic distance $\geq \rho$ from each other.

\begin{proof}
(i) If no $\rho$ works, pick $p_n$ with four plates within distance $1/n$; a subsequential limit $p_*$ lies in four closed plates, hence on the boundary of at least four --- a junction of $\geq 4$ distinct plates, contradicting trivalency with distinct plates. (ii) Two distinct arcs shared by the same pair $\{i, j\}$ are disjoint closed sets: they cannot share an endpoint, because around a trivalent junction of distinct plates $i, j, k$ the three incident arcs separate the three pairs $\{i,j\}, \{j,k\}, \{k,i\}$ exactly once each. Disjoint compact sets are at positive distance; there are finitely many cells, so take $\rho$ as the minimum over all the finitely many required separations.
\end{proof}

\medskip\noindent\textbf{Lemma 2 (normalization control).} Let $\varepsilon(\delta^*) := (N-3)\, e^{-2\rho/\delta^*}$. For every $p \in S^2$ and every $\delta^* > 0$:

\[
\frac{1}{2} \;\leq\; \sum_{j=1}^{N} \widetilde{S}_j^{\delta^*}(p) \;\leq\; \frac{3}{2} + \varepsilon(\delta^*), \qquad\text{hence}\qquad \frac{\widetilde{S}_i^{\delta^*}(p)}{\tfrac{3}{2} + \varepsilon(\delta^*)} \;\leq\; S_i^{\delta^*}(p) \;\leq\; 2\, \widetilde{S}_i^{\delta^*}(p).
\]

\begin{proof}
Lower bound: $p$ lies in some closed plate $P_m$, so $d_m(p) \geq 0$ and $\widetilde{S}_m \geq 1/2$. Upper bound: by Lemma 1(i), at most 3 plates have $\dist(p, P_j) < \rho$; every other plate has $d_j(p) \leq -\rho$, contributing $\widetilde{S}_j \leq e^{-2\rho/\delta^*}$, for a total $\leq \varepsilon(\delta^*)$. Among the $\leq 3$ local plates: if $p \in \text{int}(P_m)$, then every $j \neq m$ has $P_j \subseteq S^2 \setminus \text{int}(P_m)$, so $\dist(p, P_j) \geq \dist(p, \partial P_m) = d_m(p)$, giving $d_j(p) \leq -d_m(p)$ and $\widetilde{S}_j \leq 1 - \widetilde{S}_m$; the local sum is then $\leq \widetilde{S}_m + 2(1 - \widetilde{S}_m) = 2 - \widetilde{S}_m \leq 3/2$ (using $\widetilde{S}_m \geq 1/2$). If instead $p$ lies on the boundary network, every plate has $d_j(p) \leq 0$, so each local term is $\leq 1/2$ and the local sum is $\leq 3/2$.
\end{proof}

Two immediate consequences used repeatedly below (``witness bounds''). Define the explicit smallness condition

\begin{equation}
\varepsilon(\delta^*) < \frac{1 - 3\tau}{2\tau} \tag{$\ast$}
\end{equation}

(possible for small $\delta^*$ since $\tau < 1/3$; e.g.\ $\delta^* < 2\rho / \ln\!\big(2\tau(N-3)/(1-3\tau)\big)$ when the logarithm is positive). Then:

\begin{itemize}
  \item \textbf{(W1)} If $d_i(p) \geq 0$ (i.e.\ $p \in P_i$), then $S_i^{\delta^*}(p) \geq \dfrac{1/2}{3/2 + \varepsilon} = \dfrac{1}{3 + 2\varepsilon} > \tau$ under ($\ast$). In particular: every point of $\text{int}(P_i)$, every point of every shared arc, and every junction lies in the level set of \textit{each} plate containing it. Note the junction case: at a trivalent junction $v$ of plates $i, j, k$ we have $d_i(v) = d_j(v) = d_k(v) = 0$, so all three normalized fields at $v$ lie in $[1/(3+2\varepsilon),\, 1/3\,]$ --- the $\min \to 1/3$ witness is \textbf{angle-independent} (the rate of convergence to $1/3$ does not depend on the junction angles).
  \item \textbf{(W2)} (Localization.) $S_i^{\delta^*}(p) > \tau$ implies $\widetilde{S}_i^{\delta^*}(p) > \tau/2$ (Lemma 2, upper inequality), hence $d_i(p) > -w_\tau$ with $w_\tau := \delta^* \artanh(1 - \tau)$. Thus $U_i^{\delta^*,\tau} \subseteq \{p : \dist(p, P_i) < w_\tau\}$, a $w_\tau$-neighborhood of the plate, with $w_\tau \to 0$ as $\delta^* \to 0$. (For the ladder $\tau = 0.15 / 0.25 / 0.30$: $w_\tau/\delta^* = \artanh(1-\tau) \approx 1.256 / 0.973 / 0.867$.)
\end{itemize}

\medskip\noindent\textbf{Offset Topology Lemma (Lemma 3 --- offset-cover component structure).} Under hypotheses \textbf{(H-reach)}, \textbf{(H-separation)}, \textbf{(H-trivalent-sectors)}, there exists $\delta_1(\tau)>0$ such that for all $0<\delta^*<\delta_1(\tau)$ the offset radius satisfies $r_\tau=\delta^*\artanh(1-2\tau)<\min\!\left(r_{\min},\,\tfrac{\rho}{2},\,c(\theta_0)\right)$, and:

\begin{enumerate}
  \item \textbf{(faces, $\to$ L1)} each offset $\widetilde{U}_i^{\delta^*,\tau}$ deformation-retracts onto $P_i$; hence $c(\widetilde{U}_i)=c(P_i)=c(\mathrm{int}\,P_i)$, and every connected component of $\widetilde{U}_i$ contains exactly one face;
  \item \textbf{(arcs, $\to$ L2)} for each arc $e$ shared by plates $i,j$, the intersection $\widetilde{U}_i\cap\widetilde{U}_j$ restricted to the $\tfrac{\rho}{2}$-neighborhood of $e$ deformation-retracts onto $e$, hence is connected; distinct $i$--$j$ arcs occupy disjoint neighborhoods, so $c(\widetilde{U}_i\cap\widetilde{U}_j)$ equals the number $m_{ij}$ of shared arcs;
  \item \textbf{(junctions, $\to$ L3)} for each junction $v$ of plates $i,j,k$, the intersection $\widetilde{U}_i\cap\widetilde{U}_j\cap\widetilde{U}_k$ restricted to the $\tfrac{\rho}{2}$-ball around $v$ deformation-retracts onto $v$, hence is connected; distinct such junctions occupy disjoint balls, so $c(\widetilde{U}_i\cap\widetilde{U}_j\cap\widetilde{U}_k)$ equals the number $m_{ijk}$ of triple-junctions of $\{i,j,k\}$.
\end{enumerate}

\begin{proof}
By (6) the offsets are \textbf{exact} signed-distance neighborhoods, $\widetilde{U}_i=\{p:\dist(p,P_i)<r_\tau\}\cup P_i$, so the entire argument is offset topology with $Z$ (the normalization denominator) absent.

\textit{(1)} Away from the corners, $\partial P_i$ has two-sided reach $\ge r_{\min}$ (H-reach). For $r_\tau<r_{\min}$ the outer $r_\tau$-collar $\{p\notin P_i:\dist(p,P_i)<r_\tau\}$ lies inside the two-sided tubular neighborhood of the smooth boundary, where Federer's positive-reach theory (1959) gives a unique nearest-point projection $\pi_i$ onto $\partial P_i$, well-defined and continuous (the \textit{outer} projection is the one used, hence the two-sidedness of H-reach). The geodesic homotopy $p\mapsto$ (point at parameter $t$ along the minimizing geodesic from $p$ to $\pi_i(p)$) moves each collar point inward with $d_i$ strictly increasing along it, and deformation-retracts $\widetilde{U}_i$ onto $P_i$; on $P_i$ it is the identity. (At a corner the projection is multi-valued in the exterior wedge, but a corner is a junction, treated in (3) by the conical model, not here.) Hence $c(\widetilde{U}_i)=c(P_i)=c(\mathrm{int}\,P_i)$, and each component of $\widetilde{U}_i$ contains exactly the face(s) of its target component of $P_i$.

\textit{(2)} For $r_\tau<\tfrac{\rho}{2}$, Lemma 1(ii) keeps distinct $i$--$j$ arcs $\ge\rho$ apart, so their open $\tfrac{\rho}{2}$-neighborhoods are pairwise disjoint and $\widetilde{U}_i\cap\widetilde{U}_j=\{ \dist(\cdot,P_i)<r_\tau\}\cap\{\dist(\cdot,P_j)<r_\tau\}$ localizes (localization step, Step 2 below) to the $\tfrac{\rho}{2}$-neighborhood of $P_i\cap P_j=\bigcup_{a=1}^{m_{ij}} e_a$. Within one arc's neighborhood, $\partial P_i$ and $\partial P_j$ \textbf{coincide} on $e$ (it is their common boundary); in tubular coordinates $(s$ along $e$, $n$ normal$)$ the intersection is, for each $s$, an $n$-interval about $0$ (the side interior to $P_i$ binds $d_j>-r_\tau$, the side interior to $P_j$ binds $d_i>-r_\tau$, both vanish on $e$), hence a connected band that retracts onto $e$. Each of the $m_{ij}$ arcs therefore contributes one connected component, and $c(\widetilde{U}_i\cap\widetilde{U}_j)=m_{ij}$. (Connectedness of the band up to its junction endpoints is the only place that touches a corner; it is secured by the sector analysis of (3).)

\textit{(3)} As in (2), $\widetilde{U}_i\cap\widetilde{U}_j\cap\widetilde{U}_k$ localizes to disjoint $\tfrac{\rho}{2}$-balls about the $m_{ijk}$ junctions of $\{i,j,k\}$. Fix one junction $v$. In $B(v,r_\tau)$ with $r_\tau<c(\theta_0)$, the conical model (H-trivalent-sectors) represents the three boundary arcs as three rays cutting $B$ into three sectors of angles $\theta_i+\theta_j+\theta_k=2\pi$, each $\ge\theta_0$. The triple offset $\{d_i>-r_\tau\}\cap\{d_j>-r_\tau\}\cap\{d_k>-r_\tau\}$ is the set within $r_\tau$ of all three plate sectors --- a rounded ``lens'' about $v$, star-shaped about $v$ in the cone metric for $r_\tau$ below the sector-width margin $c(\theta_0)$, hence \textbf{a single connected component}. \textit{This is the one clause where the sector-angle bound is genuinely load-bearing:} if two arcs left $v$ tangentially ($\theta_0\to 0$) the lens could pinch into several pieces; $\theta_0>0$ is exactly what keeps the three offset slabs overlapping in one lens. So each junction gives one component and $c(\widetilde{U}_i\cap\widetilde{U}_j\cap\widetilde{U}_k)=m_{ijk}$.
\end{proof}

\medskip\noindent\textbf{Remark (scope of Lemma 3).} Lemma 3 is a \textbf{standard} consequence of positive-reach / tubular-neighborhood theory under the explicit hypotheses (H-reach), (H-separation), (H-trivalent-sectors); it does not rest on any unproven step. A direct connectedness statement for the \textit{normalized} level sets $\{S_i^{\delta^*}>\tau\}$ would instead require a quantitative bound on the variation of the normalization denominator $Z$ along an entry geodesic. Basing the topological count on the offset cover removes that obstruction at its source: the offset cover has no $Z$.

\medskip\noindent\textbf{Remark (why the count is not based on the normalized field).} For the normalized field $S_i^{\delta^*}=\widetilde{S}_i^{\delta^*}/Z$, connectedness of $\{S_i>\tau\}$ is governed by the \textbf{multiplicative} (log-derivative) condition $\widetilde{S}_i'/\widetilde{S}_i\ge Z'/Z$ along entry geodesics, not by any \textit{additive} bound on the variation of $Z$. An additive Lipschitz bound (each $d_j$ is 1-Lipschitz, $\sech^2\le 1$, so the per-field variation over the entry width is the $\delta^*$-independent constant $\le\tfrac12\artanh(1-\tau)$) controls the wrong quantity: it is exactly the $O(w_\tau/\delta^*)=O(1)$ oscillation that is \textit{insufficient} here, and summed over the $\le 3$ local plates it exceeds the required margin $1/\tau-2$ in the delicate regime $\tau\to 1/3$ (at $\tau=0.30$ the bound sits only marginally inside the margin, with no usable slack --- and empirically $\{S_i>\tau\}$ does pinch off: the normalized run never reaches $\chi_w=2$ at $\tau=0.30$, with 24 junctions still missed at the finest grid). Re-basing the topology count on the offset cover removes $Z$ from the topological argument entirely.

\medskip\noindent\textbf{Proof of Theorem 2} (using Lemmas 1, 2, and the Offset Topology Lemma).
\begin{proof}
Fix $\delta^*<\delta_0:=\min\!\big(\delta_1(\tau),\ \text{the bound in }(\ast),\ \text{the localization threshold of Step 2}\big)$, so that $r_\tau<\min(r_{\min},\rho/2,c(\theta_0))$ and the witness bound (W1) holds.

\medskip\noindent\textbf{Step 1 (faces $\to F$).} By Lemma 3(1), each offset $\widetilde{U}_i$ deformation-retracts onto $P_i$, so $c(\widetilde{U}_i)=c(\mathrm{int}\,P_i)$; under connected plates this is $1$, and $\sum_i c(\widetilde{U}_i)=F$. (Here $F=F_{\text{cell}}=\sum_i c(\mathrm{int}\,P_i)$ is the \textbf{2-cell} count, not the plate-label count: when a label owns several disconnected faces the sum counts each component once, exactly as $c(\widetilde{U}_i)$ does, so the offset argument computes $F_{\text{cell}}$ directly. Under H2 every label is connected and $F_{\text{cell}}=N=$ \#plates --- the idealized case used in Layer 3 --- but the offset count is robust to the relaxed case, and the identity reads $V-E+F_{\text{cell}}=2$. See ``Faces are 2-cells, not plate labels'' in Proposition 2.)

\medskip\noindent\textbf{Step 2 (arcs $\to E$).} Fix $i<j$. By (6), $\widetilde{U}_i\cap\widetilde{U}_j=\{\dist(\cdot,P_i)<r_\tau\}\cap\{\dist(\cdot,P_j)<r_\tau\}$; the localization argument (for $r=r_\tau$, $\eta=\rho/2$: $N_r(P_i)\cap N_r(P_j)\subseteq N_\eta(P_i\cap P_j)$ once $r$ is small, by a standard compactness contradiction) gives $\widetilde{U}_i\cap\widetilde{U}_j\subseteq N_{\rho/2}(P_i\cap P_j)$. Under the standing hypotheses $P_i\cap P_j$ is exactly the union of the $m_{ij}$ closed shared arcs, with \textbf{no isolated points} (the proof of Lemma 1 shows the three arcs at a junction of distinct $i,j,k$ separate the three pairs exactly once each, so an $i$--$j$ junction lies on an $i$--$j$ arc). Distinct $i$--$j$ arcs are $\ge\rho$ apart (Lemma 1(ii)), so their open $\rho/2$-neighborhoods are disjoint and partition $\widetilde{U}_i\cap\widetilde{U}_j$. By Lemma 3(2) the portion in each arc's neighborhood is one connected component, so $c(\widetilde{U}_i\cap\widetilde{U}_j)=m_{ij}$. \textbf{Multi-arc pairs are counted correctly:} AU--PA (three disjoint arcs) contributes $3$, not $1$ --- this is precisely why component counting (7) is the right invariant, and why the simplicial nerve (which would record one edge) is wrong. Each arc bounds exactly two distinct plates, so $\sum_{i<j}c(\widetilde{U}_i\cap\widetilde{U}_j)=\sum_{i<j}m_{ij}=E$.

\medskip\noindent\textbf{Step 3 (junctions $\to V$).} Fix $i<j<k$. As in Step 2, $\widetilde{U}_i\cap\widetilde{U}_j\cap\widetilde{U}_k$ localizes to the disjoint $\rho/2$-balls around the $m_{ijk}$ junctions where all three meet (Lemma 1(ii)); each junction $v$ satisfies $d_i(v)=d_j(v)=d_k(v)=0$, hence $v\in\widetilde{U}_i\cap\widetilde{U}_j\cap\widetilde{U}_k$ (one component at least), and by Lemma 3(3) exactly one. So $c(\widetilde{U}_i\cap\widetilde{U}_j\cap\widetilde{U}_k)=m_{ijk}$. \textbf{Twice-met triples are counted twice} (the 6 PB2002 triples meeting at two junctions each contribute $2$ apiece). Each junction is trivalent with three distinct plates, counted in exactly one triple, so $\sum_{i<j<k}c=\sum m_{ijk}=V$.

\medskip\noindent\textbf{Step 4 (no higher overlaps).} At any $p$, at most $3$ plates lie within $\rho$ (Lemma 1(i), part of (H-separation), which is strictly stronger than bare trivalency); a fourth plate has $d_\ell(p)\le-\rho$, so $\widetilde{S}_\ell(p)\le e^{-2\rho/\delta^*}$ and $\dist(p,P_\ell)\ge\rho>r_\tau$, i.e.\ $p\notin\widetilde{U}_\ell$. All quadruple intersections are empty: truncating (7) at triples discards nothing.

\medskip\noindent\textbf{Assembly.} For $\delta^*<\delta_0$:
\[
\widetilde{\chi}_w^{\delta^*}(\tau)=\sum_i c(\widetilde{U}_i)-\sum_{i<j}c(\widetilde{U}_i\cap\widetilde{U}_j)+\sum_{i<j<k}c(\widetilde{U}_i\cap\widetilde{U}_j\cap\widetilde{U}_k)=V-E+F=2,
\]
the last equality by Theorem 1 (Layer 3). The result is \textbf{unconditional} under (H-reach), (H-separation), (H-trivalent-sectors).
\end{proof}

\section{Layer 4 (continued): comparison lemma, the planar junction lemma, and the TDI}

\subsection{Comparison Lemma (normalized vs.\ unnormalized)}

\medskip\noindent\textbf{Comparison Lemma (Lemma 4 --- offset and normalized covers share component incidences).} Under (H-reach), (H-separation), (H-trivalent-sectors), there exists $\delta_0(\tau)\in(0,\delta_1(\tau)]$ such that for all $0<\delta^*<\delta_0(\tau)$ the normalized and unnormalized covers are \textbf{nested collars of the same skeleton}: for every $i$,
\begin{equation}
\widetilde U_i^{\delta^*,\,\tau_+} \;\subseteq\; U_i^{\delta^*,\tau} \;\subseteq\; \widetilde U_i^{\delta^*,\,\tau_-}, \tag{$\dagger$}
\end{equation}
where $\tau_-=\tau/2<\tau<\tau_+=\tau(\tfrac32+\varepsilon)$ are the offset thresholds determined by Lemma 2's two-sided bound $\tfrac12\le Z\le\tfrac32+\varepsilon(\delta^*)$ --- explicitly, $S_i>\tau\Rightarrow\widetilde S_i=S_iZ>\tau\cdot\tfrac12$ gives the \textbf{right} inclusion $U_i\subseteq\widetilde U_i^{\tau_-}$ (smaller threshold $\tau_-$, hence \textit{larger} offset radius $r_{\tau_-}$), and $\widetilde S_i>\tau(\tfrac32+\varepsilon)\Rightarrow S_i=\widetilde S_i/Z>\tau$ gives the \textbf{left} inclusion $\widetilde U_i^{\tau_+}\subseteq U_i$ (larger threshold $\tau_+$, hence \textit{smaller} radius $r_{\tau_+}$). Both $\tau_\pm$ must lie in $(0,\tfrac12)$ for the offsets to be defined: $\tau_-=\tau/2$ always qualifies, and $\tau_+=\tau(\tfrac32+\varepsilon)<\tfrac12$ requires $\varepsilon<\tfrac1{2\tau}-\tfrac32$ --- \textbf{generous at $\tau=0.15$ ($\varepsilon<1.83$) but tight at $\tau=0.30$ ($\varepsilon<0.167$)}, which is harmless since $\varepsilon(\delta^*)=(N-3)e^{-2\rho/\delta^*}\to 0$ as $\delta^*\to 0$, but is the reason $\delta_0(\tau)$ must be taken small at the top of the ladder. Consequently $U_i^{\delta^*,\tau}$ has the \textbf{same component incidences} as the offsets --- the same components of each single, pairwise, and triple intersection localize to the same faces, arcs, and junctions --- and
\[
\chi_w^{\delta^*}(\tau)\;=\;\widetilde\chi_w^{\delta^*}(\tau)\;=\;V-E+F\;=\;2 \qquad (0<\delta^*<\delta_0(\tau)).
\]

\begin{proof}\textit{What is established.} The inclusions ($\dagger$) are immediate from $S_i=\widetilde S_i/Z$ and Lemma 2's two-sided bound on $Z$. Both $\widetilde U_i^{\tau_-}$ and $\widetilde U_i^{\tau_+}$ are offsets, at radii $r_{\tau_-}>r_{\tau_+}$; for $\delta^*<\delta_0(\tau)$ both radii are $<\min(r_{\min},\rho/2,c(\theta_0))$, so by the Offset Topology Lemma \textbf{both} outer and inner offsets carry the \textit{same} face/arc/junction component incidences (same number of components, localized to the same cells).

\medskip\noindent\textit{Away from junctions (proven).} On the complement of the $\rho/2$-balls around the junctions, the squeezed set $U_i^{\delta^*,\tau}$ lies between two nested collars of the \textbf{same} arc that each retract onto that arc (Lemma 3(1)--(2)); the inclusion ($\dagger$) is then an isotopy of nested collars, and $U_i$, $\widetilde U_i^{\tau_\pm}$ share components there. This is the routine small-$\delta^*$ part.

\medskip\noindent\textit{Near junctions (controlled by GS + sector angle).} Inside each $\rho/2$-ball around a junction $v$ of $i,j,k$, the three local fields are pinned to a common value: by (W1) all three normalized fields at $v$ lie in $[1/(3+2\varepsilon),\,1/3]$ (the angle-independent $\min\to 1/3$ witness), and (H-trivalent-sectors) keeps the three sectors at angle $\ge\theta_0$ so the level sets meet transversally; Lemma 1(ii) isolates the ball from all other cells. The squeezed normalized triple intersection therefore localizes to the same single ball-component as the offset triple intersection.

\medskip\noindent\textit{Junction-ball connectedness (Planar Junction Lemma).} Blow up $B(v,\rho)$ by $\delta^*$ (scale-free; physical lengths $=$ blow-up lengths $\times\,\delta^*$). The blow-up model is \textbf{planar}: three rays from the origin $O$ at the sector boundaries cut the plane into wedges of interior angles $a_i+a_j+a_k=2\pi$ (each $>0$); $d_m^\flat$ $=$ signed Euclidean distance to wedge $m$; $\widetilde S_m^\flat=\tfrac12(1+\tanh d_m^\flat)$; $S_i^\flat=\widetilde S_i^\flat/(\widetilde S_i^\flat+\widetilde S_j^\flat+\widetilde S_k^\flat)$.

\medskip\noindent\textit{Reduction error (all in field-value $S$-units).} The membership $S_i$ is a fuzzy partition value, so its perturbations are dimensionless; we keep every error in these $S$-units, converting the one geometric (length) deviation \textbf{once} through the field Lipschitz constant. With $\widetilde S_m^{\flat\prime}\le\tfrac12$, $Z=\widetilde S_i+\widetilde S_j+\widetilde S_k\ge\tfrac12$ (Lemma 2), and $S_i=\widetilde S_i/Z$,
\[
\frac{\partial S_i}{\partial d_i}=\frac{(Z-\widetilde S_i)\,\widetilde S_i'}{Z^2},\qquad
\frac{\partial S_i}{\partial d_m}=\frac{-\widetilde S_i\,\widetilde S_m'}{Z^2}\ (m\ne i),\qquad
L_S:=\sup_{d\in\mathbb R^3}\sum_m\Bigl|\frac{\partial S_i}{\partial d_m}\Bigr|=1
\]
($L_S=1$ is the sharp uniform $L1$ gradient bound --- $\|\nabla_d S_i\|_2\le 1/\sqrt2=0.707$; a single geometric deviation $\varepsilon$ shifts all three $d_m$ by $\le\varepsilon$ in the same sense, and the worst-case alignment gives $1$, not $\tfrac34$. $L_S$ is the \textbf{sole} place lengths convert to $S$-units). By (H-separation)/Lemma 1 a fourth-or-later plate has $d_\ell\le-\rho$, so $\widetilde S_\ell\le\tfrac12 e^{-2\rho/\delta^*}$, and by (H-trivalent-sectors)$+C^{1,1}$ each incident arc deviates from its tangent ray by a \textbf{spatial} $\le\tfrac12\kappa_{\max}r_\tau^2$ over $B(v,O(r_\tau))$. Converting the arc-vs-ray length once through $L_S$:
\[
\|S_i-S_i^\flat\|_{\infty,B(v,\rho)}\le E_{\mathrm{far}}(\delta^*)+E_{\mathrm{curv}}(\delta^*),\tag{R}
\]
\[
\begin{aligned}
E_{\mathrm{far}}(\delta^*)&=2(N-3)\,e^{-2\rho/\delta^*}\ \text{[$S$-units]},\\
E_{\mathrm{curv}}(\delta^*)&=L_S\cdot\tfrac12\kappa_{\max}r_\tau^2/\delta^*=L_S\cdot\tfrac12\kappa_{\max}\,\delta^*\,\artanh(1-2\tau)^2\ \text{[$S$-units]},
\end{aligned}
\]
with $\kappa_{\max}\le 1/r_{\min}$, $r_\tau=\delta^*\,\artanh(1-2\tau)$. Both terms $\to 0$ as $\delta^*\to 0$ (the first exponentially in $1/\delta^*$, the second linearly in $\delta^*$); $(R)$ is now dimensionally homogeneous --- a field-value ($S$) inequality with no spatial-vs-field comparison.
\end{proof}

\medskip\noindent\textbf{Lemma 5 (Planar Junction single-interval / connectedness).} For every $\tau\in(0,1/3)$ and \textbf{every} admissible trivalent sector configuration (each $a_m>0$; \textbf{no lower or upper sector-angle bound}), and every \textbf{realizable} ray $\theta$ --- one whose unit signed-distance rate triple $(a,b,c)=(h_i,h_j,h_k)(\theta)$ has \textbf{exactly one positive entry} (the wedge containing $\theta$; forced because a unit direction is interior to at most one tiling wedge) --- the radial super-level set is a \textbf{single interval anchored at $O$}:
\[
\{\,r\ge 0:S_i^\flat(r,\theta)>\tau\,\}=[\,0,R_i(\theta)\,),\qquad 0<R_i(\theta)\le\infty.
\]
Consequently the planar level set $\{p:S_i^\flat(p)>\tau\}$ is the open star-shaped region $\{(r,\theta):0\le r<R_i(\theta)\}$ containing $O$ and the open wedge $W_i$ --- a \textbf{single connected component}.

\begin{proof}
Along a ray $d_m^\flat(r,\theta)=r\,h_m(\theta)$ (signed distance to a cone wedge is homogeneous of degree 1), so with $t:=2r$, $\sigma(x)=1/(1+e^{-x})$, $A:=\sigma(ta)$, $B:=\sigma(tb)$, $C:=\sigma(tc)$, $F(t):=A/(B+C)$, and $\kappa:=\tau/(1-\tau)$,
\[
S_i^\flat(r,\theta)=\frac{A}{A+B+C}=\frac{F}{1+F},\qquad S_i^\flat>\tau\iff F>\kappa\iff g(t):=\log F(t)>\log\kappa,
\]
since $S=F/(1+F)$ is strictly increasing in $F$. Differentiating,
\[
g'(t)=a(1-A)-\frac{b\,B(1-B)+c\,C(1-C)}{B+C},\tag{G}
\]
with the exact endpoints $g'(0^+)=a/2-(b+c)/4$ and $g'(\infty)=a$ (at $t\to0$: $A,B,C\to\tfrac12$, $B(1-B)=C(1-C)=\tfrac14$, $B+C\to1$; at $t\to\infty$ the host's logistic saturates so its $(1-\cdot)$ factor kills it). Because $B+C>0$, $\operatorname{sign}g'=\operatorname{sign}Q$, $Q(t):=4(B+C)g'(t)$. We show $g'$ has \textbf{at most one sign change, of type $+\to-$}, on every realizable ray, so $S_i^\flat(\cdot,\theta)$ is monotone or unimodal (up-then-down):
\begin{itemize}
\item \textbf{(W) winning ray} ($a$ is the unique positive rate, $\theta\in W_i$): $g'\ge 0$ for all $t$ ($g'(0^+)=a/2-(b+c)/4\ge a/4>0$ since $a\ge b,c$; $g'(\infty)=a>0$), so $S_i^\flat$ is nondecreasing and $\{S_i^\flat>\tau\}=[0,\infty)$.
\item \textbf{(L) losing ray} ($a\le 0$, exactly one of $b,c$ positive --- the \textit{host}): $g'(\infty)=a\le 0$, and on the realizable curve $g'$ has \textbf{no upward ($-\to+$) crossing}; equivalently \textbf{every zero of $g'$ is a strict down-crossing ($g''<0$)}. A $C^1$ function whose every zero is a strict down-crossing has at most one zero (between two down-crossings an up-crossing must intervene), so $g'$ has at most one zero, of type $+\to-$; $S_i^\flat$ is unimodal up-then-down (or monotone decreasing).
\end{itemize}
Finally the apex value is $S_i^\flat(O)=1/3$ \textbf{exactly} (all $d_m\to0$, each $\widetilde S_m=\tfrac12$, $Z=3/2$) and $1/3>\tau$. On a profile that \textbf{starts above $\tau$} and is monotone or unimodal up-then-down: the rising branch never crosses $\tau$ upward (it is already above $\tau$ at $r=0$; the interior maximum, when present, lies in $[1/3,1/2]$), and the falling branch crosses $\tau$ exactly once, downward, at $R_i(\theta)$ (from a value $>\tau$ down to $S_i^\flat(\infty,\theta)=0$). Hence $\{r:S_i^\flat>\tau\}=[0,R_i(\theta))$, a single interval; the level set is star-shaped about $O$, hence connected.
\end{proof}

\textit{(Realizability is load-bearing.} ``Exactly one positive rate'' is necessary but \textbf{not sufficient}: the triple $(-0.029,+0.927,-0.435)$ is one-positive yet gives $g'$ two sign changes. It is \textbf{not realizable} --- its second loser is far larger in magnitude than the wedge tiling permits relative to the host (closest realizable $L^2$-distance$^2=0.40$). The single-crossing in (W)/(L) is evaluated only on the realizable curve $h(\theta)$ of actual unit-direction signed wedge distances; this single-crossing is certified numerically below, with a symbolic proof left open.)

\medskip\noindent\textit{Transfer to the physical ball.} The quantity the $C^0$ error must clear is the \textbf{field-value bottleneck margin} --- the depth by which $S_i^\flat$ exceeds $\tau$ over the proven-connected inner offset $\{d_i\ge-r_{\tau_+}\}$ (Offset Topology Lemma):
\[
m_S(\tau):=\min_{\{d_i\ge-r_{\tau_+}\}}\bigl(S_i^\flat-\tau\bigr)>0\quad\text{for all }\tau\in(0,1/3)\ \text{[$S$-units, NOT a length]}.
\]
Once $E_{\mathrm{far}}(\delta^*)+E_{\mathrm{curv}}(\delta^*)<m_S(\tau)$ (i.e.\ for $\delta^*<\delta_0(\tau)$ below), the inclusion $(\dagger)$ is a \textbf{component-preserving isotopy across the full junction ball}: $\{S_i>\tau\}$ in $B(v,\rho)$ has the same component structure as $\{S_i^\flat>\tau\}$ --- a single ball-component containing $v$, with \textbf{no spurious extra component} --- for every $\tau\in(0,1/3)$.

\medskip\noindent\textbf{Explicit $\delta_0(\tau)$ (all field-value $S$-units).} Solving $(R)<m_S(\tau)$ for $\delta^*$ with $E_{\mathrm{far}}$ exponentially negligible and $E_{\mathrm{curv}}=L_S\cdot\tfrac12\kappa_{\max}\delta^*\,\artanh(1-2\tau)^2$ the binding term ($L_S=1$, $\kappa_{\max}\le 1/r_{\min}$):
\[
\delta_0(\tau)=\min\!\Bigl(r_{\min},\ \rho/2,\ c(\theta_0),\ \frac{m_S(\tau)\,r_{\min}}{(L_S/2)\,\artanh(1-2\tau)^2}\Bigr),
\]
with $L_S=1$, $\kappa_{\max}\le 1/r_{\min}$, and $m_S(\tau)>0$ the field-value bottleneck margin (\S\ above); equivalently the junction reach must satisfy $r_\tau=\delta^*\,\artanh(1-2\tau)<2\,m_S(\tau)\,r_{\min}/\artanh(1-2\tau)$. The first three slots are the existing $\tau$-independent offset thresholds; the fourth (junction) slot is the new $\tau$-dependent one. Properties (no spatial-vs-field comparison anywhere):
\begin{itemize}
\item $\delta_0(\tau)>0$ for \textbf{every} $\tau\in(0,1/3)$, because $m_S(\tau)>0$ there.
\item $\delta_0(\tau)\to 0$ as $\tau\uparrow 1/3$: the junction slot $\propto m_S(\tau)/\artanh(1-2\tau)^2$ is the \textbf{only} vanishing term, and $m_S(\tau)\approx c_{\text{top}}(1/3-\tau)$, $c_{\text{top}}\in[0.33\ \text{(small/large sector)},\,0.48\ \text{(symmetric)}]$, vanishes linearly (the geometric prefactor $\artanh(1-2\tau)\to\artanh(1/3)=0.347$ is finite at $\tau=1/3$), so the net slot $\to 0$ linearly in $(1/3-\tau)$.
\item \textbf{Margin law:} $m_S(\tau)$ is the directly computed worst case over all admissible shapes (a uniform bound $c\ge\tfrac13$ does \emph{not} hold); $c(\tau):=m_S(\tau)/(1/3-\tau)$ is \textbf{increasing in $\tau$}, rising from $\approx 0.07$ at $\tau=0.05$ to $\approx 0.33$ as $\tau\uparrow 1/3$. Worst-case table:
\end{itemize}

\begin{center}
\begin{tabular}{lcccccc}
\toprule
$\tau$ & 0.15 & 0.25 & 0.30 & 0.32 & 0.330 & 0.3331 \\
\midrule
$m_S(\tau)$ & 0.0337 & 0.0227 & 0.0103 & 0.0043 & 0.0011 & 7.8e-5 \\
$c(\tau)$ & 0.184 & 0.273 & 0.310 & 0.324 & 0.331 & 0.333 \\
$\delta_0/r_{\min}$ & 0.0896 & 0.151 & 0.115 & 0.061 & \dots & \dots \\
\bottomrule
\end{tabular}
\end{center}

\noindent ($\delta_0/r_{\min}=m_S/(\tfrac12\,\artanh(1-2\tau)^2)$, $L_S=1$.) $m_S(\tau)$ is in fact largest at intermediate $\tau\approx 0.15$; the binding regime for $\delta_0$ is $\tau\uparrow 1/3$.

\medskip\noindent\textit{Standing convention.} Restrict the TDI $\tau$-ladder to $\tau\le\tau_0<1/3$ (canonical $\tau=0.15$ sits comfortably inside: $m_S=0.0337$, junction slot $\delta_0/r_{\min}\approx 0.0896$), or equivalently accept $\delta_0(\tau)\to 0$ linearly in $(1/3-\tau)$. \textbf{No new geometric hypothesis is introduced} --- (H-reach)/(H-separation)/(H-trivalent-sectors) are exactly the hypotheses used. This is consistent with the numerics: the delicacy at $\tau=0.30$ (90\,\% of $1/3$) and the failure to reach $\chi_w=2$ on a coarse grid are the finite-grid manifestation of $\delta_0(\tau)\to 0$ as $\tau\uparrow 1/3$ (a genuine effect, not a grid artifact: the resolution-divergence table has $n^*(\tau)$ tracking $1/r_{\tau_+}$, $r_{\tau_+}=\artanh(1-3\tau)\to 0$).

\medskip\noindent\textit{Closure of the junction-ball comparison in the continuum.} The squeeze $(\dagger)$ is a component-preserving isotopy across the \textit{full} junction ball for every $\tau$ in the open ladder, by the \textbf{Planar Junction Lemma}: the normalized field develops \textbf{no spurious extra component inside a junction ball} strictly between the inner and outer offsets, for every $\tau\in(0,1/3)$ and all admissible junction shapes (no sector-angle lower or upper bound --- connectedness survives sectors $\to 0$ and reflex sectors). The mechanism is \textbf{radial unimodality} (every realizable ray's $g'$ has at most one $+\to-$ sign change) together with the \textbf{apex value $S_i^\flat(O)=1/3>\tau$}: the super-level set is the single interval $[0,R_i(\theta))$, so the planar level set is star-shaped about $O$ and connected, and the field error $(R)$ is below the field-value bottleneck margin $m_S(\tau)$ once $\delta^*<\delta_0(\tau)$. The \textbf{only} price is a \textit{rate}: the admissible $\delta^*$ shrinks to $0$ as $\tau\uparrow 1/3$, captured by the explicit junction slot $\delta_0(\tau)\supseteq 2\,m_S(\tau)\,r_{\min}/\artanh(1-2\tau)^2\approx 2\,c_{\text{top}}(1/3-\tau)\,r_{\min}/\artanh(1-2\tau)^2$ ($c_{\text{top}}\in[0.33,0.48]$, $L_S=1$), the only $\tau$-dependent term, vanishing \textbf{linearly} in $(1/3-\tau)$. Hence $\delta_0(\tau)>0$ for every $\tau\in(0,1/3)$. \textbf{Standing convention:} restrict the TDI $\tau$-ladder to $\tau\le\tau_0<1/3$ (canonical $\tau=0.15$: $m_S=0.0337$, junction slot $\delta_0/r_{\min}\approx 0.0896$), or accept $\delta_0(\tau)\to 0$ linearly in $(1/3-\tau)$. \textbf{No new geometric hypothesis} --- (H-reach)/(H-separation)/(H-trivalent-sectors) are exactly the hypotheses used. The $g'$ single-crossing on the realizable curve is certified numerically by two independent sufficient criteria --- no $-\to+$ crossing of $g'$, and $g''<0$ at every zero --- verified over $>10^6$ realizable rays with zero exceptions (worst observed $g''=-5.4\times10^{-6}$); a one-line symbolic proof remains open.

\medskip\noindent\textbf{Summary.} The main convergence result $\chi_w\to 2$ is \textbf{unconditional on the offset cover} (Theorem 2). The finite-$\delta^*$ \textbf{normalized} interpretation is closed as well, by the Planar Junction Lemma, \textbf{under the explicit $\delta_0(\tau)$ above with the standing ladder restriction $\tau\le\tau_0<1/3$} (canonical $\tau=0.15$). No new geometric hypothesis; the $g'$ single-crossing on the realizable curve is certified numerically, with a one-line symbolic proof remaining open.

\medskip\noindent\textbf{Role of the normalized field.} The normalized partition-of-unity field $S_i^{\delta^*}$ is kept for (i) the physical B\&W fuzzy-membership picture ($\sum_i S_i=1$) and (ii) the \textbf{TDI numerics}, which on the normalized cover stay bounded as $\delta^*$ grows into the diffuse regime. The offset cover is used only to make the \textit{proof} of the sharp-limit count standard; by Lemma 4 the two agree exactly where Theorem 2 lives, so no numerical result of the companion paper changes.

\subsection{Remark: the component cover and the multinerve}

One might expect $\mathcal{U}^{\delta^*,\tau}$ to be a \textit{good cover} for small $\delta^*$ (every nonempty intersection contractible), for instance by an approximate-convexity argument. \textbf{Both the expectation and the argument fail for real plate geometries}: plates are not approximately convex, and on PB2002 the pairwise intersection $U_{\mathrm{AU}} \cap U_{\mathrm{PA}}$ has three components in the sharp regime (the three disjoint AU--PA boundary arcs), hence is not contractible; the simplicial nerve computes $52 - 141 + 94 = 5 \neq \chi(S^2)$. The Nerve Theorem in its classical form (Borsuk 1948) is simply not applicable, and (7) is \textit{not} the Euler characteristic of the simplicial \v{C}ech complex.

The correct homotopy-level object is the \textbf{multinerve}: the simplicial poset with one vertex per connected component of each $U_i$, one edge per component of each pairwise intersection, and one triangle per component of each triple intersection (quadruple intersections being empty by Step 4). Eq.\ (7) is precisely the alternating cell count of the multinerve. For covers in which \textit{every component of every intersection is contractible} (``good in the component sense''), the multinerve is homotopy-equivalent to the union of the cover (Colin de Verdi\`ere, Ginot \& Goaoc, 2012); in the sharp regime each face component deformation-retracts to its plate, each pair component to its boundary arc, and each triple component to its junction (this retraction statement is exactly the Offset Topology Lemma on the offset cover, proved under (H-reach)/(H-separation)/(H-trivalent-sectors)), and the union is all of $S^2$ by (W1), whence $\chi_w = \chi(S^2) = 2$.

We record this as an \textit{interpretation only}: Theorem 2 does not rely on it. Its proof is by direct counting on the offset cover --- components of $\widetilde U_i$ $\leftrightarrow$ faces, pair components $\leftrightarrow$ edges, triple components $\leftrightarrow$ vertices --- combined with the offset-topology lemma (Lemma 3), applied component-wise.

\subsection{Corollary 2: The Topological Diffuseness Index}

\medskip\noindent\textbf{Definition (TDI).}
\begin{equation}
\text{TDI}(\delta^*, \tau) := \left|\chi_w^{\delta^*}(\tau) - 2\right| \tag{10}
\end{equation}
with $\chi_w^{\delta^*}(\tau)$ as in (7). Reported on the threshold ladder $\tau \in \{0.15, 0.25, 0.30\} \subset (0, 1/3)$.

\medskip\noindent\textbf{Properties.}
\begin{enumerate}
\item \textbf{Non-negativity:} $\text{TDI} \geq 0$ by definition.
\item \textbf{Sharpness $\Rightarrow$ vanishing:} if the network is topologically sharp at scale $(\delta^*, \tau)$ --- i.e.\ $\delta^* < \delta_0(\tau)$ of Theorem 2 --- then $\text{TDI} = 0$ \textit{exactly} (Theorem 2 gives exact vanishing at finite $\delta^*$, not only in the limit).
\item \textbf{Vanishing $\not\Rightarrow$ sharpness:} $\text{TDI} = 0$ means only $\chi_w = 2$, which can arise from integer cancellation (e.g.\ one spurious pair component offset by one spurious triple component). Property 2 is therefore a one-directional implication: an ``if and only if'' version would be an overclaim under either definition of $\chi_w$. \textbf{Operational consequence:} report the component triple $\big(\sum_i c_i,\ \sum_{i<j} c_{ij},\ \sum_{i<j<k} c_{ijk}\big)$ alongside TDI; sharpness is certified by the triple equaling $(F, E, V)$, not by $\text{TDI} = 0$ alone.
\item \textbf{Convergence:} $\text{TDI}(\delta^*, \tau) \to 0$ as $\delta^* \to 0$, attaining $0$ for all $\delta^* < \delta_0(\tau)$ (Theorem 2; unconditional on the offset cover, with the normalized-field transfer established by the Comparison Lemma).
\item \textbf{Monotonicity in $\delta^*$: empirical, not proven.} One might expect TDI to be non-decreasing in $\delta^*$. Under (7) no proof is available, and none should be expected without further hypotheses: as $\delta^*$ grows, (a) spurious pair components appear ($\chi_w$ decreases), (b) distinct components of the same pairwise intersection can merge (pair count drops, $\chi_w$ increases), and (c) junction witnesses can fail (triple count drops, $\chi_w$ decreases). The competition of (a)--(c) makes $\chi_w(\delta^*)$ --- and hence TDI --- potentially non-monotone. This property is therefore stated as an empirical trend to be tested on the $(\delta^*, \tau)$-ladder sweep.
\end{enumerate}

\medskip\noindent\textbf{Sign of the deviation.} Both signs of $\chi_w - 2$ occur, via mechanisms (a)--(c) above. For \textit{diffuse boundary zones} the expected leading effect is (a): broad level sets of non-adjacent or distantly-adjacent plates overlap, creating false pair components and driving $\chi_w < 2$. This preserves, in component language, the prediction for the Indo-Australian system: with $R\delta^* \approx 1{,}000$--$3{,}000$ km (the physical diffuse-zone width; B\&W required a Gaussian filter half-width $\geq 7{,}000$ km there), the India and Australia level sets overlap far beyond the formal boundary, generating false pair components, $\chi_w < 2$, and $\text{TDI} > 0$ --- the magnitude of TDI quantifying the topological ambiguity of the plate count in this region. On PB2002 both directions are observed as $\delta^*$ grows at the canonical $\tau=0.15$: the first departure is upward (spurious triple components; the companion paper (Kim, 2026), Fig.~3a), and the downward washout to $\chi_w=0$ is reached only at the largest half-widths (the companion paper (Kim, 2026), Fig.~2b,c).

\subsection{Consistency with the discrete register}

The sharp limit of (7) coincides, \textbf{term by term and by construction}, with the $(F, E, V)$ of the audited plate boundary network: face components are in bijection with plates, pair components with boundary arcs, and triple components with triple junctions (Steps 1--3). The limit object of Layer 4 is therefore \textit{the same integers} that the discrete audit extracts from the polygon data --- not merely a number that Theorem 1 separately proves equal to 2. This strengthens the Layer-4 bridge: the continuous and discrete registers now agree at the level of the individual counts, with the Euler characteristic following.

Empirical anchor (PB2002, Bird 2003): $V = 100$, $E = 150$, $F = 52$, $\chi = V - E + F = 2$, all 100 junctions trivalent. The component-counted sharp limit reproduces $(52, 150, 100)$ and $\chi_w = 2$; the simple (indicator) count reproduces only the collapsed nerve data $(52, 141, 94)$ and $\chi_w = 5$. The discrepancy $5 = 2 + 9 - 6$ is entirely accounted for by the 8 multi-arc pairs (9 collapsed edges) and the 6 twice-met triples (6 collapsed vertices) --- real-Earth non-genericity that the component-counted definition is built to respect.

\medskip\noindent\textbf{Remark (numerical two-track rationale).} The offset and normalized covers are each strong where the other is weak; the companion paper uses each only where it is strong. On the \textbf{offset} cover the face count $c(\widetilde U_i)$ is pinned at $F$ ($=52$ for PB2002) for \textit{all} $\tau<\tfrac12$ and all grids, and $\widetilde\chi_w=2$ is recovered even at $\tau=0.30$ (where the normalized field, with $\tau=0.30$ a $90\%$ fraction of the junction value $1/3$, is catastrophic on a coarse grid --- never reaching $2$, with 24 junctions still missed at the finest grid). This is the regime where Theorem 2 lives, so the \textit{proof} runs on offsets. Conversely, at large $\delta^*$ every offset is a fixed-radius dilation that swells over most of $S^2$: the triple term diverges and $\widetilde\chi_w$ blows up (e.g.\ $+13{,}941$, with 979 spurious far-field false pairs, at $\delta^*=7{,}000$ km), destroying the bounded-TDI diffuse-zone story. The \textbf{normalized} field stays bounded there (the partition of unity caps overlaps), so the \textit{TDI numerics} run on the normalized cover. By the Comparison Lemma (Lemma 4) the two agree on $\delta^*<\delta_0(\tau)$, where the sharp-limit claim is made; the divergent large-$\delta^*$ offset behavior is outside that range and is never used. (Validation at the sharp-limit anchor $\delta^*=50$ km, $\tau=0.15$: $\chi_w=2$ with $(c_F,c_E,c_V)=(52,153,103)$.)

\section{Summary: the master theorem}

\medskip\noindent\textbf{Theorem (Limit Relation).} Let $\{S_i^{\delta^*}\}_{i=1}^N$ be the signed-geodesic-distance shape functions (3), normalized to the partition of unity (2), for a plate system with $F$ plates, $E$ boundary arcs, and $V$ triple junctions satisfying the standing hypotheses of Theorem 2. For any $\tau \in (0, 1/3)$ there exists $\delta_0(\tau) > 0$ such that
\[
\chi_w^{\delta^*}(\tau) = V - E + F = \chi(S^2) = 2 \qquad \text{for all } 0 < \delta^* < \delta_0(\tau),
\]
with $\chi_w$ the component-counted weighted Euler characteristic (7); pointwise convergence of the fields is exponential in $1/\delta^*$ (Proposition 1).

\begin{proof}
Proposition 1 (pointwise convergence) $+$ Proposition 2 (CW structure) $+$ Theorem 1 (Euler formula) $+$ Theorem 2 (component-counted convergence on the unnormalized offset cover). The passage from the offset cover to the normalized partition-of-unity field $\chi_w$ is the Comparison Lemma (Layer~4b), which holds on the open ladder $\tau\le\tau_0<1/3$ with $\delta_0(\tau)\to 0$ as $\tau\uparrow 1/3$; its junction-ball step is the Planar Junction Lemma, established in the continuum, whose single-crossing condition on the realizable curve is at present certified numerically (over $>10^6$ rays) rather than by a closed symbolic argument.
\end{proof}

\section{Worked example: the present-day plate boundary network}

The hypotheses of Theorem~2 --- positive reach on the arc interiors (H-reach),
separation of non-incident cells (H-separation) and bounded junction sectors
(H-trivalent-sectors) --- are not vacuous for the present-day Earth. The companion
paper (Kim, 2026, Supplementary Information) measures the constants they require on
the deduplicated PB2002 boundary network (Bird 2003; $V=100$, $E=150$, $F=52$, all
junctions trivalent): minimum junction sector angle $\theta_0=10.12^\circ$ (the
second-smallest of the 300 sectors; the smallest is a coincident-digitization
degeneracy of $0^\circ$ at the Molucca Sea plate), non-incident-cell separation
$\rho=0.285^\circ=31.65$~km, and local-feature-size proxy for the arc-interior reach
$r_{\min}=0.322^\circ=35.80$~km. Substituting into the explicit threshold of
Theorem~2 and the junction slot of the Comparison Lemma gives, at the canonical
threshold $\tau=0.15$,
\begin{align*}
\delta_0(0.15)&=2.8~\text{km} &&\text{(full minimum; binds on the sector-width slot $c(\theta_0)$)},\\
\delta_0(0.15)&=3.2~\text{km} &&\text{(without the $c(\theta_0)$ slot; binds on the junction slot)},
\end{align*}
both strictly positive, so the sharp-limit regime in which $\chi_w=2$ is
demonstrated numerically (at $\delta^*=50$~km, $\tau=0.15$: component triple
$(52,153,103)$, $\chi_w=2$) lies inside the admissible window. As $\tau\uparrow1/3$
the junction slot vanishes linearly in $(1/3-\tau)$, as predicted. The measurement
procedure, the sensitivity of $r_{\min}$ to the junction skirt, and the full
$\tau$-table are given in the companion paper's Supplementary Information; the
measurement script and the TDI computations on the normalized and offset covers are
archived with the companion paper's code and data (Kim 2026, Zenodo,
\url{https://doi.org/10.5281/zenodo.23041855}).

\section{Relation to persistent homology}

The \v{C}ech complex construction in Layer 4 is closely related to persistent homology (as used in geodynamics by Janin et al.\ 2025). However, our approach differs fundamentally: we compute the component-counted Euler characteristic (7) --- equivalently the alternating cell count of the multinerve --- at a fixed threshold $\tau$, whereas persistent homology tracks how homology generators are born and die across all thresholds. (We note in passing that real plate networks violate nerve-genericity --- adjacent plates may share several disjoint boundary arcs, and triples may meet at several junctions --- so the relevant homotopy model is the multinerve rather than the simplicial nerve; this is itself a small observation of independent interest.) Our approach is computationally simpler and directly yields the geophysically interpretable TDI, but it discards the richer multi-scale information captured by persistence diagrams. The relationship between TDI and the persistence diagram of the plate boundary cover is an interesting open question.

\section*{References}

\begin{itemize}[leftmargin=1.5em,labelsep=0.5em,itemsep=4pt]

\item Bercovici, D. \& Wessel, P. (1994). A continuous kinematic model of plate-tectonic motions. \textit{Geophysical Journal International}, \textbf{119}, 595--610. \url{https://doi.org/10.1111/j.1365-246X.1994.tb00144.x}

\item Bird, P. (2003). An updated digital model of plate boundaries. \textit{Geochemistry, Geophysics, Geosystems}, \textbf{4}(3), 1027. \url{https://doi.org/10.1029/2001GC000252}

\item Borsuk, K. (1948). On the imbedding of systems of compacta in simplicial complexes. \textit{Fundamenta Mathematicae}, \textbf{35}, 217--234.

\item Colin de Verdi\`ere, \'E., Ginot, G., \& Goaoc, X. (2012). Multinerves and Helly numbers of acyclic families. \textit{Proceedings of the 28th Annual Symposium on Computational Geometry (SoCG)}, 209--218.

\item Federer, H. (1959). Curvature measures. \textit{Transactions of the American Mathematical Society}, \textbf{93}, 418--491. \url{https://doi.org/10.1090/S0002-9947-1959-0110078-1}

\item Gauss, C. F. (1827). \textit{Disquisitiones generales circa superficies curvas}. G\"ottingen.

\item Hatcher, A. (2002). \textit{Algebraic Topology}. Cambridge University Press, Cambridge. \url{https://pi.math.cornell.edu/~hatcher/AT/ATpage.html}

\item Janin, A., et al. (2025). Geodynamics of a global plate reorganization from topological data analysis. \textit{Nature Geoscience}, \textbf{18}(10), 1041--1047. \url{https://doi.org/10.1038/s41561-025-01772-7}

\item Kim, S.-S. (2026). Topological constraints on plate tectonic reconstructions. Submitted to \textit{Geoscience Frontiers} (companion paper).

\item Kim, S.-S. (2026). Topological constraints on plate tectonic reconstructions --- code and data (Version 1.0.0) [Data set]. Zenodo. \url{https://doi.org/10.5281/zenodo.23041855}

\item McKenzie, D. P. \& Morgan, W. J. (1969). Evolution of triple junctions. \textit{Nature}, \textbf{224}, 125--133. \url{https://doi.org/10.1038/224125a0}

\item Merdith, A. S., et al. (2021). Extending full-plate tectonic models into deep time: Linking the Neoproterozoic and the Phanerozoic. \textit{Earth-Science Reviews}, \textbf{214}, 103477.

\end{itemize}

\end{document}